# Probing the formation of femtosecond laser-induced periodic surface structures on silica films by ultrafast small-angle X-ray scattering

*Dominik Kaczmarek[#], Jean-Luc Déziel, Thies Johannes Albert, Matthias Weise, Jerzy Antonowicz, Robert Carley, Devesh Chopra, Loïc Le Guyader, Laurent Mercadier[❀], Giuseppe Mercurio, Roman Minikayev, Mianzhen Mo, Andreas Scherz, Ryszard Sobierajski, Yanwen Sun, Martin Teichmann, Peter Zalden, Klaus Sokolowski-Tinten, and Jörn Bonse[*]*

D. Kaczmarek, Faculty of Physics and Center for Nanointegration Duisburg-Essen (CENIDE), University of Duisburg-Essen, Lotharstraße 1, 47057 Duisburg, Germany, ORCID: https://orcid.org/0009-0006-2810-7274
[#] dominik.kaczmarek@uni-due.de

J.-L. Déziel, Bentley Systems, Incorporated, 685 Stockton Drive, Exton, PA 19341, USA, ORCID: https://orcid.org/0000-0002-2522-1665

T.J. Albert, Faculty of Physics and Center for Nanointegration Duisburg-Essen (CENIDE), University of Duisburg-Essen, Lotharstraße 1, 47057 Duisburg, Germany;
European XFEL, Holzkoppel 4, 22869 Schenefeld, Germany, ORCID: https://orcid.org/0000-0002-8117-9334

M. Weise, Bundesanstalt für Materialforschung und -prüfung (BAM), Unter den Eichen 87, 12205 Berlin, Germany

J. Antonowicz, Faculty of Physics, Warsaw University of Technology, Koszykowa 75, 00-662 Warsaw, Poland, ORCID: https://orcid.org/0000-0002-7781-7540

R. Carley, European XFEL, Holzkoppel 4, 22869 Schenefeld, Germany, ORCID: https://orcid.org/0000-0002-0015-7089

D. Chopra, European XFEL, Holzkoppel 4, 22869 Schenefeld, Germany, ORCID: https://orcid.org/0009-0006-6987-2044

L. Le Guyader, European XFEL, Holzkoppel 4, 22869 Schenefeld, Germany, ORCID: https://orcid.org/0000-0002-5731-3724

L. Mercadier, European XFEL, Holzkoppel 4, 22869 Schenefeld, Germany, ORCID: https://orcid.org/0000-0003-0606-0461
[❀] present address: IRFM (Institute for Magnetic Fusion Research), CEA-Cadarache, Saint-Paul-lez-Durance, 13108 France

G. Mercurio, European XFEL, Holzkoppel 4, 22869 Schenefeld, Germany, ORCID: https://orcid.org/0000-0003-4221-6670

R. Minikayev, Institute of Physics, Polish Academy of Sciences, Aleja Lotnikow 32/46, 02668 Warsaw, Poland, ORCID: https://orcid.org/0000-0002-4334-7472

M. Mo, Energy Sciences Directorate, SLAC National Accelerator Laboratory, 2575 Sand Hill Rd, Menlo Park, CA 94025, USA, ORCID: https://orcid.org/0000-0002-2962-0815

A. Scherz, European XFEL, Holzkoppel 4, 22869 Schenefeld, Germany, ORCID: https://orcid.org/0000-0001-8187-7178

R. Sobierajski, Institute of Physics, Polish Academy of Sciences, Aleja Lotnikow 32/46, 02668 Warsaw, Poland, ORCID: https://orcid.org/0000-0003-2580-6900

Y. Sun, Linac Coherent Light Source, SLAC National Accelerator Laboratory, 2575 Sand Hill Rd, Menlo Park, CA 94025, USA, ORCID: https://orcid.org/0000-0001-5926-6565

M. Teichmann, European XFEL, Holzkoppel 4, 22869 Schenefeld, Germany, ORCID: https://orcid.org/0000-0003-4215-9792

P. Zalden, European XFEL, Holzkoppel 4, 22869 Schenefeld, Germany, ORCID: https://orcid.org/0000-0002-8530-0576

K. Sokolowski-Tinten, Faculty of Physics and Center for Nanointegration Duisburg-Essen (CENIDE), University of Duisburg-Essen, Lotharstraße 1, 47057 Duisburg, Germany, ORCID: https://orcid.org/0000-0002-7979-5357

J. Bonse, Bundesanstalt für Materialforschung und -prüfung (BAM), Unter den Eichen 87, 12205 Berlin, Germany, ORCID: https://orcid.org/0000-0003-4984-3896
* joern.bonse@bam.de

**Abstract**

Ultrashort X-ray pulses, as provided by X-ray Free Electron Lasers (XFELs), offer unique opportunities to probe the formation of laser-induced nanostructures with sizes even below the optical diffraction limit, with sub-µm and sub-ps spatial and temporal resolution. Using ultrafast small-angle X-ray scattering (SAXS), we probe the formation of nanoscatterers and nanocracks in reciprocal space as they evolve into different types of laser-induced periodic surface structures (LIPSS) on silica films. Our focus is on the dynamics and evolution of nanoscatterers and high-spatial-frequency-LIPSS (HSFL), with spatial periods below 100-nm. Our experiments include an *in-situ* analysis of the pulse-by-pulse evolution of HSFL at a fixed sample location, as well as fs time-resolved delay and fluence scans using an ultrafast *pump-probe* scheme. We find that nanoscatterers and periodic nanocracks form at delay times of hundreds of picoseconds and subsequently evolve into HSFL. At low laser fluences, inter-pulse incubation effects become important during multi-pulse irradiation. Simultaneously, inter-pulse feedback leads to an ordering and improve the regularity of the HSFL. The experiments are complemented by theoretical analyses, including numerical Finite-Difference Time-Domain (FDTD) simulations. These combined analyses allow us to propose a five-stage model of LIPSS formation on silica films upon irradiation with fs-laser pulses.

## 1. Introduction

The processing of *laser-induced periodic surface structures* (LIPSS) is a simple and robust one-step strategy for surface functionalization of nearly any material through quasi-periodic, sub-micrometer, grating-like surface topographies [1-3]. These topographies allow for a wide range of applications in optics, medicine, and tribology, e.g., for anti-fogging, anti-icing, and anti-reflective surfaces [4,5], product marking of precious goods via LIPSS patterns as security tags [6,7], improved tissue cell-growth for medical implants [8,9], antibacterial surfaces [10], energy savings through the reduction of friction and wear [11], magnetic flux manipulation on superconducting thin films [12], etc.

LIPSS usually form electromagnetically seeded in a self-ordered way through inter-pulse feedback upon multi-pulse irradiation in the focal region of a focused laser beam. They typically exhibit a clear correlation to the linear laser beam polarization, i.e., either parallel or perpendicular to it. LIPSS are classified according to their spatial periods and orientations relative to the laser beam polarization as near-wavelength sized *low spatial frequency LIPSS* (LSFL) or deep-sub-wavelength sized *high spatial frequency LIPSS* (HSFL). Depending on the irradiated type of material that rules the specific formation mechanism and the historic order of their description in the pertinent literature, both types of LIPSS can be further classified into two subcategories (as type I or type II) [13].

Since the turn of the millennium, HSFL have particularly gained attention due to their nanometric feature size far below the optical diffraction limit [14]. These small-scale structures can be generated solely with ultrashort laser pulses with durations in the ps- to fs-range. For dielectrics, HSFL form predominantly for laser photon energies below the band gap of the irradiated material. These structures are classified as HSFL-I and originate from pulse-by-pulse, feedback-enforced, collective optical near-field scattering effects at laser-excited near-surface defects [15].

Due to their small spatial periods, $\Lambda_{HSFL}$, of less than a few hundred nanometers, HSFL are inaccessible through optical microscopy. Thus, *ex-situ* scanning electron microscopy (SEM) or atomic force microscopy (AFM) is usually employed to visualize and further characterize HSFL in post-irradiation experiments. However, these time-demanding approaches typically prevent studying the pulse-by-pulse evolution of HSFL *in-situ* at a specific (fixed) micrometric surface spot. Instead, experiments are performed with different pulse numbers applied to different sample locations, i.e., to statistically similar surface structures.

*In-situ* analysis of the HSFL in reciprocal space using a scattered or diffracted probe beam represents a straightforward solution for monitoring their pulse-by-pulse evolution at a fixed sample position. While this can be easily implemented with optical radiation for the near-wavelength-sized LSFL [16-18], the small-scale HSFL require a shorter probing wavelength in the XUV or X-ray spectral range. Thus, *Synchrotron* or *Free Electron Laser* radiation is required for studying the evolution of HSFL in a site-selective manner [19], e.g., via *Small-Angle X-ray Scattering* (SAXS) [20].

SAXS primarily probes the lateral changes of the X-ray refractive index, which is essentially determined by laser-induced mass density variations in the sample, e.g., via phase transitions (melting, evaporation) or material displacement (sound, cracking, ablation), and is not directly affected by intra-pulse laser-induced conduction/valence band electron density changes or inter-pulse electronic defect formation. Note that the scattered intensity patterns can be affected by both the geometrical shape of microscopically scattering objects and, for ensembles of multiple objects, also by their correlated spatial arrangement [20].

The potential of superior spatial resolution through X-rays probe wavelengths was recognized by a group of researchers around S. Nolte et al., who probed through *ex-situ* SAXS experiments the properties of so-called “nanogratings”, previously prepared by a tightly focused fs-laser beam in the bulk of fused silica or borosilicate glasses [21-23]. Apart from the nanogratings, the authors were able to resolve the presence of several tens of nm sized voids (pores), acting as laser-induced scattering centers, and further evolving into sheets of nanocracks during the nanograting formation. In the context of LIPSS, Rebollar et al. exploited the capabilities of X-rays for probing *in-situ* the pulse number dependent evolution of UV ns-laser generated LSFL-II with 150 nm to 250 nm spatial period on polymer film surfaces via *grazing-incidence small-angle X-ray scattering* (GISAXS) for pulse numbers between 25 and 2000 [24]. Sokokolowski-Tinten et al. explored the formation dynamics of LSFL-I on thin silicon films upon single-pulse irradiation in time-resolved ps-optical pump / fs-XUV probe scattering experiments, suggesting the additional presence of capillary waves with periods around 150 nm at the surface of transiently molten silicon films [25,26]. Ksenzov et al. used multiple ultrashort XUV-pulses in a diffuse scattering geometry to probe the fs-laser pulse excitation of surface acoustic waves (SAWs) that are predominantly coupled at LSFL-I spatial frequencies of the surface roughness on thin metal films for laser fluences well below the damage threshold of the irradiated material [27].

In this study, we go beyond these previous works and utilize the high brightness of ultrashort X-ray pulses generated by X-ray free-electron lasers (XFELs) to investigate the pulse-by-pulse evolution of LIPSS with a size of about 100 nm, generated by fs laser pulses on thin silicon dioxide ($SiO_2$) films, using SAXS in a transmission geometry. These *in-situ* pulse-by-pulse experiments enable to explore the LIPSS regularity through ordering inter-pulse feedback effects and are complemented by time- and fluence-resolved single-pulse pump-probe SAXS experiments, as well as theoretical modelling of the electromagnetic absorption in the laser-irradiated silica. This joint approach allows us to reveal the formation dynamics of laser-induced nanoscatterers, to determine the formation time of HSFL-induced nanocracks and to identify different stages of HSFL and LSFL formation on thin dielectric films.

## 2. Time-Resolved In-Situ Small-Angle X-ray Scattering

The in-situ SAXS experiments were performed at the soft X-ray SCS (Spectroscopy and Coherent Scattering) instrument [28] of the European XFEL (EuXFEL, Schenefeld, Germany). 200 nm thick $SiO_2$ layers were deposited by physical vapor deposition on commercially available sample carriers, consisting of 300 × 300 µm$^2$ sized 300 nm thick amorphous $Si_3N_4$ membrane windows arranged in large arrays and supported by a silicon wafer frame (cf. **Figure 1**). The selected thicknesses of the $SiO_2$ layer and $Si_3N_4$ membrane ensured sufficient transmission for the chosen X-ray photon energy. For more details on the sample preparation the reader is referred to Sect. 5.1 in the Materials and Methods part below.

Figure 1 shows a scheme of the femtosecond optical pump / SAXS probe experiment. The $SiO_2$ layers were irradiated at an angle of incidence of 11.5° to the sample surface normal, either by a single pulse or by a sequence of multiple ($N$) fs-laser pulses at 400 nm wavelength (3.1 eV photon energy, < 50 fs pulse duration) [29]. The pump laser beam was focused onto the center of a selected sample window to a 1/e$^2$ beam diameter of $2w_0 \approx 90$ µm. The peak laser fluence in front of the sample was controlled by a half-wave plate polarizer combination and set to values between 0.05 J cm$^{-2}$ and 1.95 J cm$^{-2}$ in the range from below to above the single-pulse laser damage threshold of the $SiO_2$ layer. An additional half-wave plate in the pump laser beam path adjusted the laser beam polarization to be approximately parallel with respect to the plane of incidence (p-pol.).

The XFEL was operated in single-pulse mode, delivering soft X-ray pulses on demand with a wavelength of $\lambda_{X\text{-ray}} = 0.83$ nm (1.5 keV photon energy; below the Si L-edge), ≈ 25 fs pulse duration. The X-ray probe beam was focused at normal incidence onto the center of the optically excited region with a spot size of about 40 µm (1/e$^2$ decay diameter of intensity). The scattered X-ray radiation was detected in transmission geometry on an area detector (4.2 Megapixel CCD camera [30], operated with 2×2 binned pixels). The detector was covered by a thin film aluminum filter to suppress residual optical radiation. To protect the CCD camera, a disc-shaped beam stop was placed at a suitable distance to block the non-scattered (direct) probe beam radiation. The sample-CCD distance determined the range of accessible spatial frequencies of the X-ray scattering vector, $q = |\boldsymbol{q}| = |\boldsymbol{k}_{X\text{-ray}} - \boldsymbol{k}_{0,X\text{-ray}}| = 4\cdot\pi\cdot\sin(\vartheta)/\lambda_{X\text{-ray}}$, with $\vartheta$ being half the X-ray scattering angle. The minimum and maximum detectable momentum transfer, $q_x$ and $q_y$, ranged from 0.00078 Å$^{-1}$ (limit of the beam stop) to 0.0242 Å$^{-1}$ (limit of the CCD detector size; corner position), corresponding to spatial dimensions between ≈ 805 nm and ≈ 26 nm, respectively. Finally, the X-ray pulse energy was adjusted (about tens of µJ per pulse) by a gas attenuator to prevent saturation of the detector while optimizing the dynamic range of the recorded single-pulse SAXS patterns. During the measurement campaign, the upper left quadrant of the CCD camera could not be properly initialized and thus did not provide relevant scattering signals.

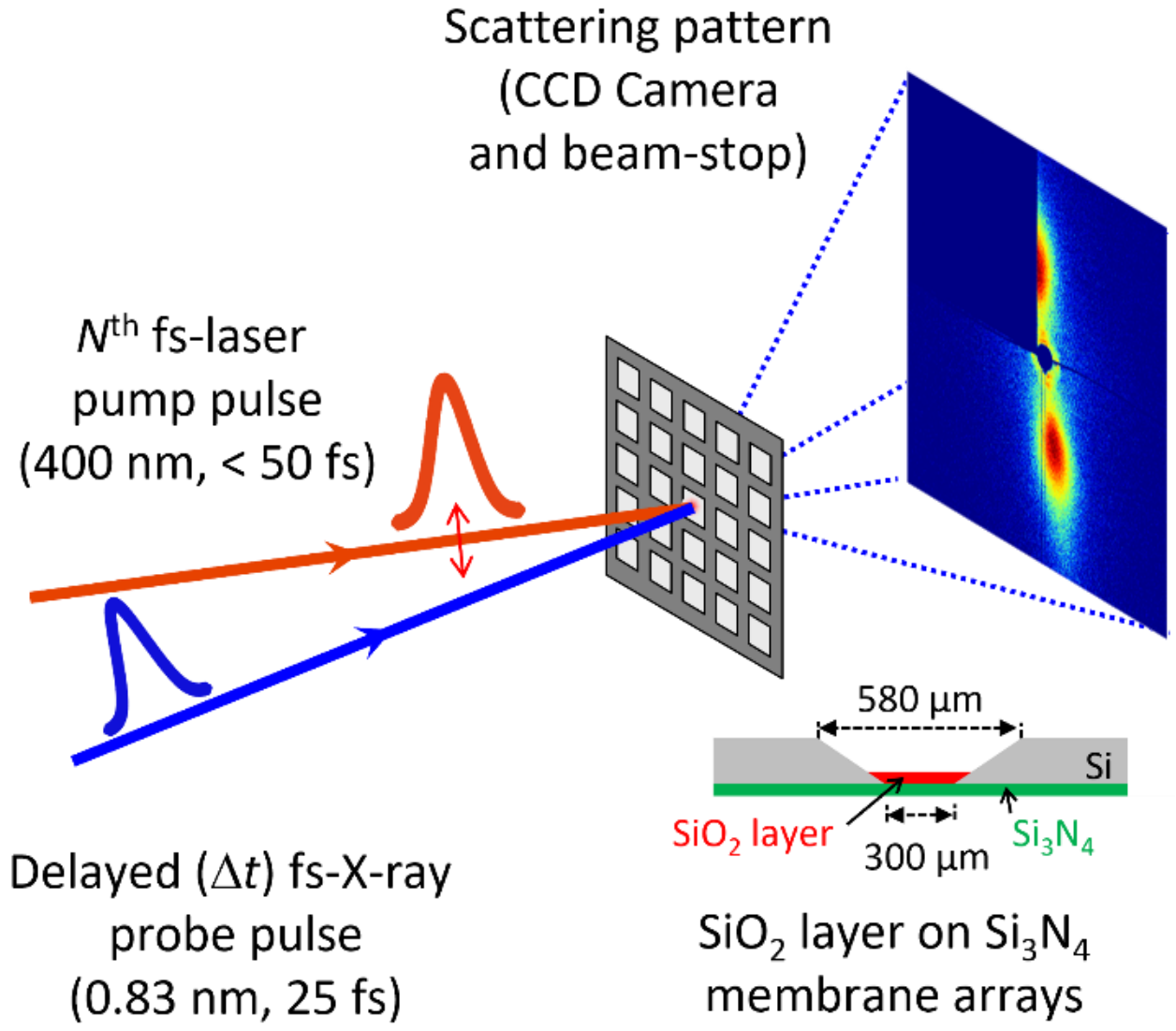

Figure 1: Experimental scheme of the time-resolved / in-situ femtosecond optical pump / small-angle X-ray scattering probe setup in transmission geometry. The red double-arrow displays the linear laser beam polarization direction. Additionally, a cross-sectional stratigraphy of a single sample window with a $Si_3N_4$ carrier membrane (green) and the covering $SiO_2$ layer (red) is drawn, where the laser/X-ray beams are incident from the top.

The laser pulses were temporally synchronized with the arrival of the XFEL pulses at the sample surface, allowing to control the inter-pulse delay time $\Delta t$ via a motorized mechanical delay stage in the range up to a few nanoseconds. Positive delay values indicate that the X-ray pulse arrives after the laser pulse at the sample. If desired, motorized translation stages enabled the sample to be moved to a pristine (non-irradiated) window. For experimental fluence- or delay-scans, the sample was always moved to a fresh sample site between the individual laser/X-ray exposures. In pump-probe experimental runs, six identical event repetitions were recorded for each delay or fluence value.

For pulse-by-pulse measurements, a single XFEL pulse probed the sample approx. 2 s after each individual fs-laser pulse. In this case, the X-ray fluence was set far below the multi-pulse damage threshold of the sample at the X-ray wavelength to prevent any modification of the sample by the probe pulse itself. In this way, a sequence of scattering patterns was recorded as a function of the number of laser pulses, $N$. The corresponding experimental pulse-by-pulse runs were terminated manually, e.g., after damage of the membrane window was observed in the SAXS patterns.

## 3. Results and Discussions

### *3.1 In-Situ Pulse-by-Pulse Small-Angle X-ray Scattering of LIPSS Formation*

Post-processing of the experimental data obtained from the in-situ pulse-by-pulse measurements was performed. for details see Sect. 5.2. The recorded X-ray flux value was used to globally normalize the intensities of the SAXS images, thereby compensating for fluctuations in the probe beam. In addition, averaged scattering images recorded before irradiation with the first laser pulse were subtracted from all images, which enabled the reduction of the X-ray induced background and isolation of the laser-induced changes. Finally, the SAXS images were smoothed by a Gaussian filter that averaged over 2.4 pixels in both the $q_x$- and $q_y$-directions.

**Figure 2** shows a sequence of SAXS patterns for 15 different numbers of laser pulses, ranging from $N = 1$ to $N = 270$, as indicated in the lower left corner of each pattern. All images are displayed in a joint color scale for the scattered intensity. The double-arrow in the frame for $N = 1$ indicates the projection of the p-polarized pump laser beam polarization onto the sample surface.

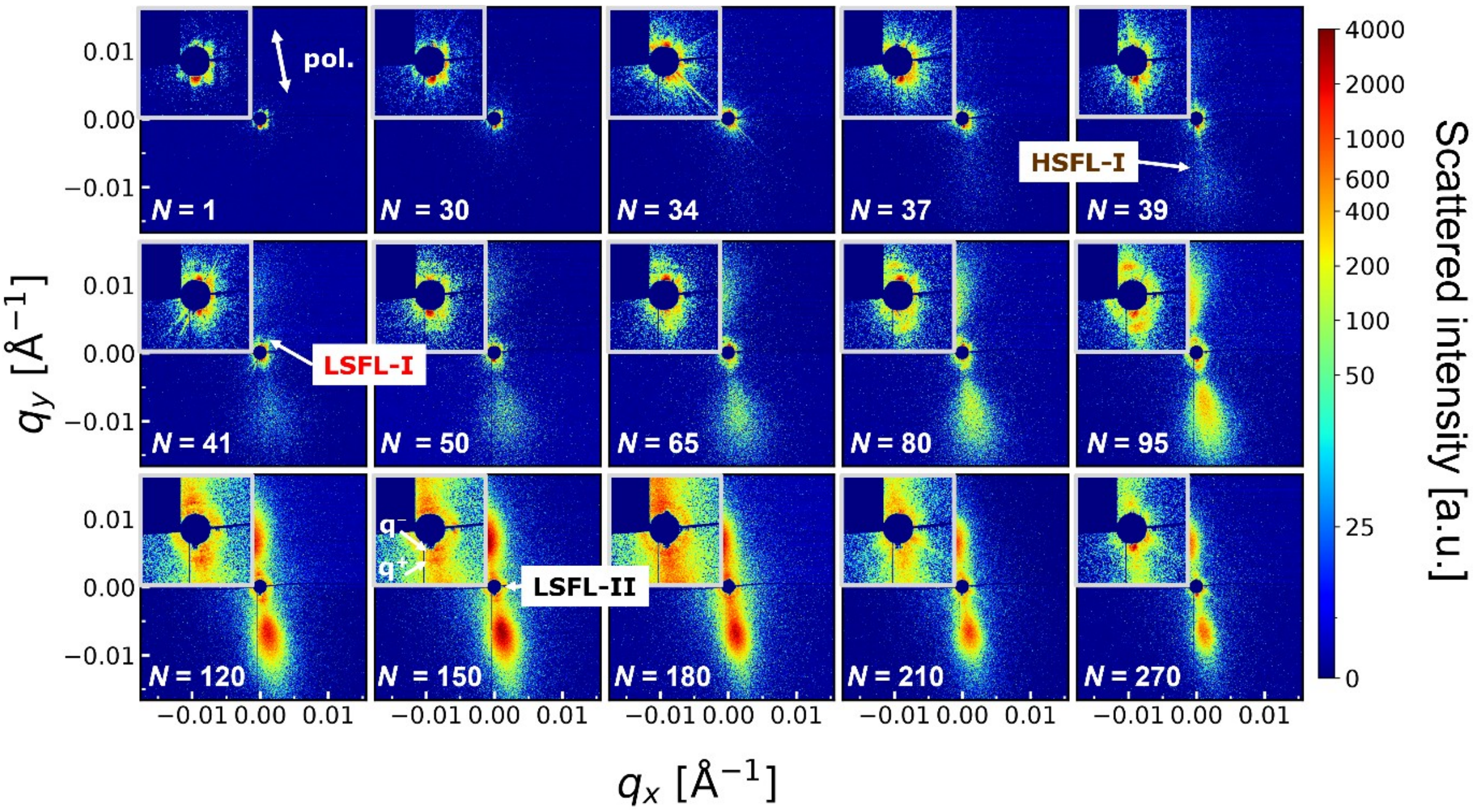

Figure 2: Selected SAXS patterns after exposure to different numbers of laser pulses, *N*, applied to the same sample spot at a fixed peak fluence of $\phi_0 = 0.27$ J cm$^{-2}$. All images are encoded in a joint color scale for the scattered intensity. The white double-arrow in the frame for *N* = 1 displays the projection of the linear laser beam polarization direction onto the sample surface. The insets show details around the beam stop in the $|q_{x,y}| \leq 0.0036$ Å$^{-1}$ range.

For a low number of laser pulses, the SAXS pattern is dominated by a signal around the beam stop (see $N = 1$). This represents leakage of the probe beam around the beam stop, resulting from imperfect background subtraction. After application of $N = 30$ laser pulses to the same spot, additional low-intensity diffuse scattering appears around the beam stop, being indicative of the onset of laser-induced modifications of the sample. For $N = 34$, some additional radial streaks can be seen, supposedly resulting from crack formation, along with a diffuse extension of the scattering pattern in the direction of the laser polarization. This becomes more pronounced after $N = 37$ laser pulses, when a clear spread of the scattered radiation appears as a symmetric double-lobe feature along the direction of the linear laser beam polarization. With just a few more laser pulses, this characteristic scattering pattern intensifies. As will be discussed later, this scattering feature can be associated with the formation of HSFL-I on the silica layer (as indicated in the frame of $N = 39$), exhibiting a LIPSS ridge orientation perpendicular to the laser beam polarization. With an increasing number of laser pulses, the intensity of this HSFL-I scattering pattern becomes significantly stronger, before starting to saturate around $N = 150$, and dropping again with larger pulse numbers up to $N = 270$. Interestingly, after an initial formation stage ($N < 95$), the center of the HSFL-I feature moderately moves towards smaller $q$-values (spatial frequencies) as $N$ increases.

After approximately $N = 40$ to 50 laser pulses, another distinct scattering feature starts to appear at smaller $q$ (close to the position of the beam stop), again in the direction of the laser polarization, representing LSFL-I (see Sect. 3.3). This scattering feature exhibits a characteristic double-arc structure that becomes more pronounced and visible with a larger number of laser pulses, see the SAXS pattern for $N = 150$. It can be associated with two different modes of LSFL-I, labeled "$q^+$" and "$q^-$", (see Sect. 3.3). For $N > 100$, another (weak) scattering feature can be observed perpendicular to the laser beam polarization (LIPSS ridges parallel to it). This scattering signature (as marked for $N = 150$) features the characteristics of LSFL-II, which are frequently observed on bulk $SiO_2$ [15,31]. Depending on the laser wavelength, LSFL-II are seeded several tens to a few hundreds of nanometers below the sample surface via radiative scattering of surface defects and intra-pulse interference with the laser pulse in the transition zone between near-field and far-fields during its propagation through the transparent sample [15,32]. This results in an accumulation of defects due to multi-pulse incubation effects within the $SiO_2$ material at localized interference maxima. The effect even manifests in thin transparent films in the presence of a reflecting interface to the substrate material [33]. In such a scenario, the LSFL-II signatures may become visible in the SAXS through structural material modifications, or when the covering part of the $SiO_2$ film material starts to be removed from the sample surface via film ablation. For pulse numbers exceeding $N \approx 165$, the LSFL-II signatures are no longer detectable in the SAXS patterns.

Another independent strong indication for the presence of laser-induced incubation effects is provided through additional pulse-by-pulse SAXS experiments performed at larger laser fluences (cf. scattering patterns in the *Supporting Information* in **Figures S1** and **S2**). From these experiments, it is evident that the number of laser pulses required to initiate film modifications and to observe HSFL-I signatures is reduced for increasing laser fluences.

To quantitatively describe the pulse-by-pulse evolution of the LIPSS characteristics, the recorded SAXS image frames were azimuthally integrated over limited angular ($\varphi$) ranges around the polarization direction ($\Delta\varphi = \pm 25°$) and in the direction perpendicular to it ($\Delta\varphi = \pm 20°$), marked by the red and green lines in **Figure 3**(a). Examples of the resulting scattering profiles, $I(q)$ ($q$: momentum transfer), are shown in Figure 3(b) as red and green curves, respectively. These experimental data were then decomposed through a least-squares fit into skewed Gaussian peaks representing the individual scattering signatures of HSFL-I, LSFL-I ($q^+$- and $q^-$-contributions) and LSFL-II, as well as into an exponentially decaying background function for considering the leakage around the beam stop.

The decomposed peaks allow to deduce the relevant parameters that quantitatively characterize the HSFL and LSFL evolution:

(i) The most frequent LIPSS period $\Lambda = \frac{2\pi}{q_{\mathrm{peak}}}$ with $q_{\mathrm{peak}}$ being the peak position of the corresponding LIPSS feature,
(ii) the relative spread/distribution of LIPSS periods $\frac{\Delta\Lambda}{\Lambda} = \frac{\Delta q_{\mathrm{FWHM}}}{q_{\mathrm{peak}}}$ with $\Delta q_{\mathrm{FWHM}}$ being the full width at half maximum (FWHM) of the scattering peaks, and
(iii) the peak intensity $I_{\mathrm{peak}}$.

Figure 3(c) assembles pulse number dependent curves of the peak scattering intensity (top panel), the most frequent LIPSS period (middle panel), and the relative spread of the LIPSS periods as a measure of their intrinsic irregularity (bottom panel) for the characteristic features of the SAXS patterns provided in Figure 2. Data for the different types of LIPSS are displayed as full violet squares (HSFL-I), open red circles (LSFL-I, $q^+$-mode, $\Lambda^+$), full blue triangles (LSFL-I, $q^-$-mode, $\Lambda^-$), and open black star symbols (LSFL-II), respectively. Additional horizontal lines in the middle panel indicate theoretical predictions for the different types of LIPSS at the given irradiation conditions. Vertical dotted lines across all panels help to distinguish different phases of film damage and LIPSS evolution. Data for the LSFL-I $q^-$-mode and the LSFL-II relative spread of periods and intensity are not provided in the top and bottom panels, since in the SAXS scattering patterns, these characteristics emerging at low spatial frequencies could not be reliably discriminated against leakage of the probe beam radiation around the beam stop.

The intensity of the HSFL-I features starts to increase weakly after $N = 37$ pulses are applied to the irradiated spot, up to a pulse number of $N \approx 80$ (top). Within this range, the HSFL-I exhibit a constant period around 80 nm (middle), with a large relative spread up to $\Delta\Lambda/\Lambda \approx 3$ (bottom). For $N > 80$, the intensity rises steeply by a factor of more than 20, reaching maximum visibility at $N = 162$ (top). During this stage, the most frequent HSFL-I period increases and saturates around ≈ 95 nm (middle panel), while its spread reduces by a factor of ≈4 and saturates at a minimum (bottom). Hence, the number of HSFL-I ridges and the regularity of their arrangement increase at the center of the laser-irradiated spot. These measurements represent the very first *in-situ* observation of the ordering of sub-100 nm HSFL-II at a fixed surface location due to positive inter-pulse feedback effects. The most frequent HSFL-I period is significantly smaller (≈ 40%) than the theoretically expected (single-pulse) value of $\Lambda_{\mathrm{HSFL\text{-}I}} \approx \lambda/(2{\cdot}n_0) = 136$ nm (horizontal dashed line), although this value always falls in the broad spread of periods. After $N = 164$, the HSFL-I intensity drops discontinuously by ≈ 50%, suggesting that the $SiO_2$ layer carrying the HSFL signatures was removed upon ablation. This trend of sudden ablation events reducing the feature's intensity also manifests at $N = 192$ and smoothly continues up to the maximum number of $N = 274$ laser pulses (bottom).

The two modes of the LSFL-I emerge after $N \approx 40$ laser pulses. The intensity of the $q^+$-mode also shows a peaked progression with few discontinuities (top). The discontinuities occur at different values of *N,* compared to the HSFL-I, suggesting that the regions of LIPSS formation may differ. The most frequent periods of the two LSFL-I modes convincingly match the values of a simple *Surface Electromagnetic Wave* (SEW) scattering model [26,34] for non-normal ($\theta = 11.5°$) incident $\lambda = 400$ nm laser radiation predicted as $\Lambda^{+/-} = \lambda/[1 \pm \sin(\theta)]$ at $\Lambda^+ = 334$ and $\Lambda^- = 500$ nm, respectively (middle, see the horizontal dash-dotted lines), indicating that the scattered wave wavelength is very close or equal to the laser wavelength, see Sect. 3.3.1. The corresponding experimental data of the relative spread $\Delta\Lambda/\Lambda$ of the $q^+$-mode largely scatter between 0.25 and 0.75, with an increasing trend with $N$ (bottom). Nevertheless, these results show that for pulse numbers up to $N \approx 100$, the intrinsic irregularity ($\Delta\Lambda/\Lambda$) is significantly smaller for the LSFL than for the HSFL – in agreement with HSFL observed on metals [35].

LSFL-II scattering signatures in the orthogonal direction (ridges oriented parallel to the laser beam polarization) manifest only in a narrow pulse number range, between $N = 120$ and 165 laser pulses (middle). Their most frequent period is rather constant around 395 nm, which exceeds the theoretically predicted (single-pulse) value of $\Lambda_{\mathrm{LSFL\text{-}II}} \approx \lambda/n_0 = 272$ nm by ≈ 45%. The systematic theoretical overestimation of the HSFL-I periods (40%) and the simultaneous underestimation of the LSFL-II periods (45%) — both are supposed to exhibit an inversely linear relationship with the refractive index of the irradiated material — therefore suggest different reasons for these discrepancies in the LIPSS periods.

One possible explanation is inter-pulse feedback through incubation effects, which decrease the optical refractive index of the material. This may result from the pulse-by-pulse generation of metastable or permanent electronic defect states in the bandgap of $SiO_2$ (e.g., self-trapped excitons or colour centers) being the origin for the required incubation stage of initial 40 to 50 laser pulses. The gradually increasing number of defects facilitates the excitation of the silica material for the subsequent laser pulses [16,31]. Moreover, each laser pulse transiently generates conduction band electrons via nonlinear absorption that largely affect the refractive index as well via their collective Drude response [31].

In our case of $SiO_2$ and at a laser wavelength of 400 nm, such a Drude model (see Section 3.3.1) predicts that, as the (intra-pulse) laser-induced conduction band electron density increases, the refractive index of the material will initially decrease by ≈ 50% from $n_0$ before rising sharply once the material turns transiently metallic. The increased LSFL-II periods observed in our pulse-by-pulse experiments, thus, suggest that a temporarily decreased refractive index (compared to $n_0$) during the laser pulse plays a role in the optical seeding phase of these structures. However, this transient decrease in the refractive index can therefore not be the origin of the simultaneously decreased experimental HSFL-I periods.

These observations are consistent with a scenario in which the LSFL-II and HSFL-I are seeded at different times (and different depths) during the laser pulse. The rising (low-intensity) part of the laser pulse can penetrate deeper into the $SiO_2$ film material (simultaneously reducing its refractive index locally via electron generation in the conduction band and seeding the LSFL-II below the surface via interference with the radiation scattered at the surface), while the subsequent high-intensity part of the laser pulse can turn the material towards a metallic state (simultaneously increasing its refractive index locally and seeding the HSFL-I at the surface through collective coherent near-field scattering).

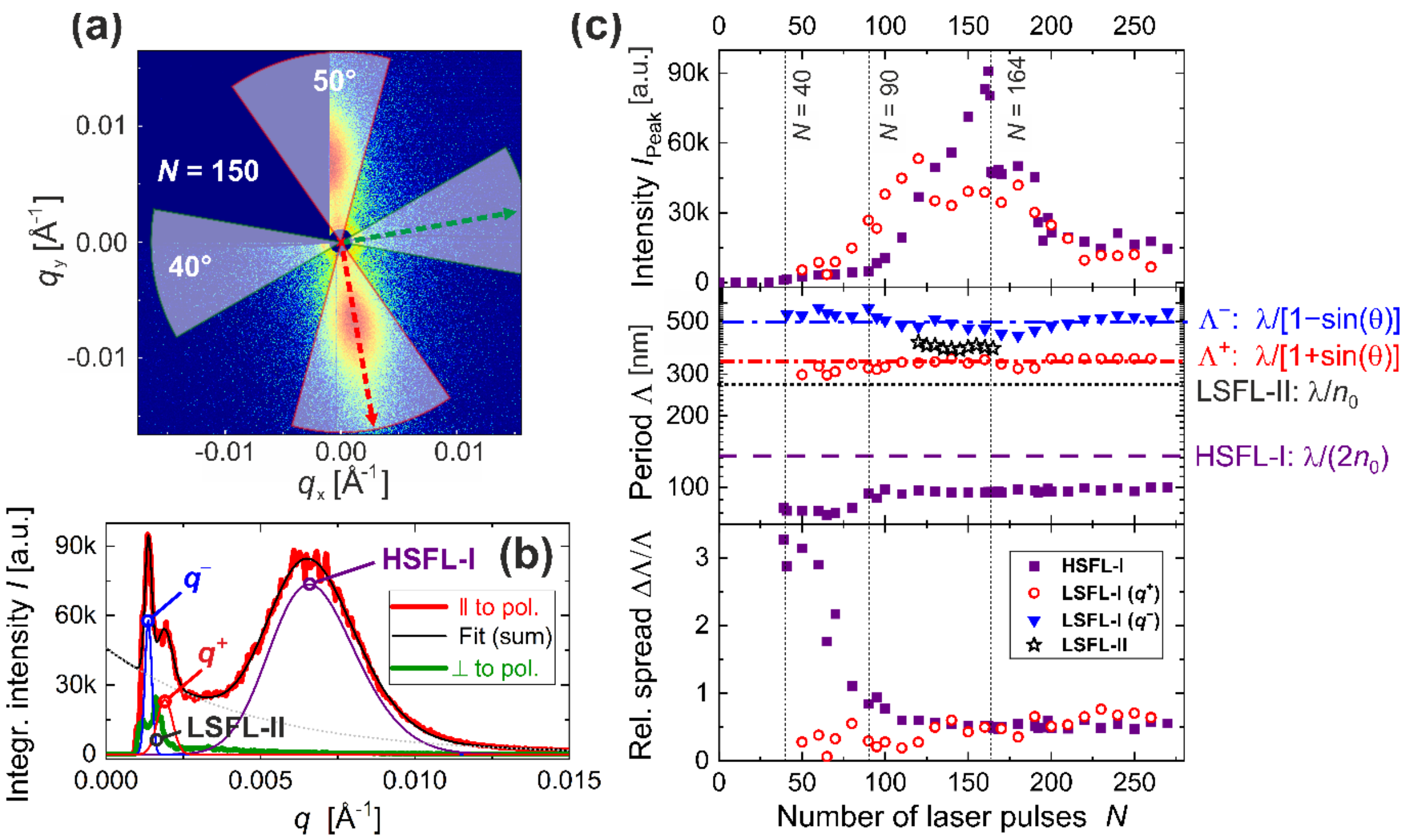

Figure 3: Characteristics of LIPSS signatures (HSFL-I, LSFL-I, LSFL-II) as a function of the number of laser pulses *N,* with a fixed peak fluence, $\phi_0$, of 0.27 J cm$^{-2}$, as derived by azimuthal integration over the corresponding LIPSS signatures in the SAXS patterns presented in Figure 2. (a) Visualization of the azimuthal integration ranges for the SAXS pattern recorded after *N* = 150 laser pulses. (b) Resulting radial intensity profiles along (red curve) and perpendicular (green curve) to the laser beam polarization direction. The characteristic peaks of the different LIPSS features are decomposed through a least-squares-fit (black curve) and exemplary marked. For more details see the text. (c) Peak scattering intensity $I_{Peak}$ (top panel), most frequent LIPSS period Λ (middle panel), and relative spread of LIPSS periods ΔΛ/Λ (bottom panel) as obtained from datadecomposition. The horizontal lines visualize theoretical predictions of single-pulse irradiation models (see the text).

To verify the persistence of the LIPSS and to clarify whether they were formed at the surface of the $SiO_2$ layer or on the $Si_3N_4$ membrane, an inspection of the sample was performed by SEM immediately after the irradiation. **Figure 4**(a) displays an overview image of the irradiated central region of the corresponding sample window. Regions featuring three different gray levels can be distinguished that can be associated with the intact $SiO_2$ layer surface (bright gray level), a partially laser ablated $SiO_2$ surface (medium gray level), and a central region, where the $SiO_2$ film is completely removed from the $Si_3N_4$ membrane (dark gray level). The presence of several cracks in the $SiO_2$ layer is evident. Higher magnification SEM imaging (Figure 4(b)) confirmed the presence of HSFL-I perpendicular to the laser beam polarization at the laser ablated $SiO_2$ layer surface. Moreover, the high-resolution inspection revealed that in the fully ablated regions no LIPSS are present on the uncovered $Si_3N_4$ membrane, neither any surface damage of the membrane (data not shown here). A measurement of the HSFL-I period averaged over 10 periods confirmed a spatial period of ≈100 nm, in excellent agreement with the SAXS data presented in Figure 3. The inset in Figure 4(b) complements a two-dimensional *Fast Fourier Transform* (2D-FFT) obtained using the free scanning probe microscopy software *Gwyddion* (vers. 2.68) [36] from the corresponding SEM micrograph. This 2D-FFT image is fully consistent with the final SAXS pattern presented in Figure 2 ($N$ = 270).

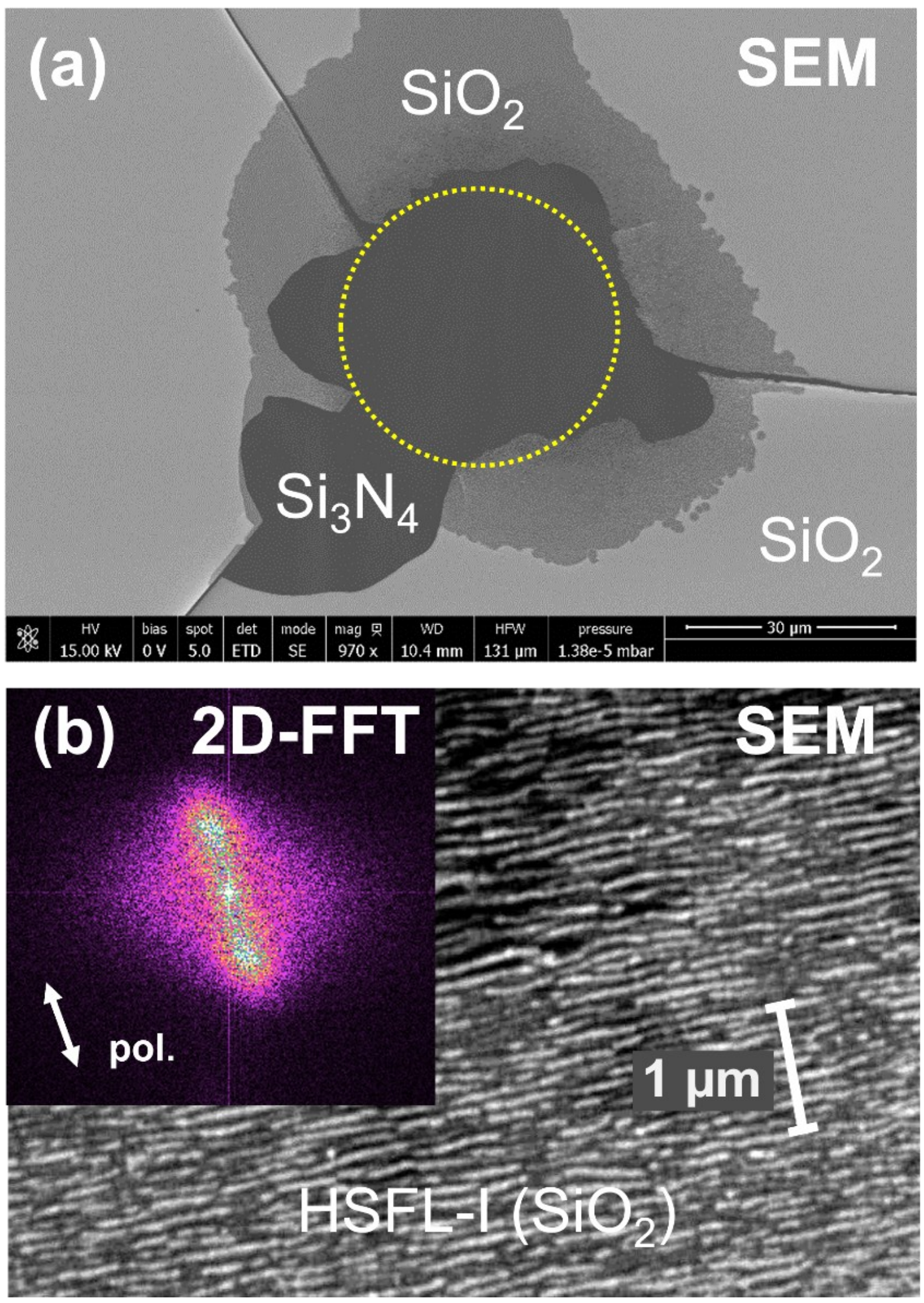

Figure 4: Top-view SEM micrographs of the final (permanent) surface modification after $N$ = 274 laser pulses at the in-situ probed sample surface previously analyzed in Figure 2. (a) Overview of the central sample window area. (b) High-resolution detail. The inset in (b) shows the corresponding two-dimensional Fast Fourier Transform (2D-FFT). The dashed yellow circle in (a) marks the $1/e^2$ beam decay diameter of the probing X-ray beam. The white double-arrow in the inset of (b) displays the projection of the direction of the linear laser beam polarization onto the sample surface.

The complete removal of the $SiO_2$ layer in the central irradiated spot area supports the interpretation provided in the discussion of Figure 3 that the discontinuous drop of the LIPSS-related SAXS-intensity is caused by laser ablation. While for the initial laser pulses, the LIPSS are predominantly formed and probed in the high-intensity center of the laser irradiated region, for a large number of laser pulses (after film ablation), the LIPSS are probed outside of the ablated spot center through the low-intensity wing of the X-ray pulses.

### *3.2 Single-Pulse Ultrafast Time-resolved Small-Angle X-ray Scattering of Fs-Laser-Excited $SiO_2$ Layers*

The above spotted discrepancy of −40% and +45% between the theoretically predicted single-pulse values and experimentally observed multi-pulse spatial periods of the HSFL-I and LSFL-II deserves further investigation. Supposedly, these differences are caused by intra- and inter-pulse feedback effects manifesting upon irradiation with multiple laser pulses. This may occur in the ablative regime via an interplay of pulse-by-pulse accumulating absorbing defects and changing surface topographies, along with the intra-pulse transient changes of the optical properties, which reinforces specific spatial periods of the grating-like LIPSS. To test this hypothesis, additional single-pulse experiments were conducted via pump-probe time-resolved SAXS.

To obtain an increased scattering signal level in the single-pulse pump-probe fs-SAXS experiments, the flux of the XFEL probe pulses was increased by a factor of ten compared to the conditions used in the pulse-by-pulse experiments discussed above, relying for the time-resolved measurements on the principle "scatter/diffract before destroy" [37]. To avoid any multi-pulse damage accumulation/incubation effects in the sample material, the sample was always moved to a fresh irradiation site (window) between the individual pump-probe events of the experimental runs (delay- or fluence-scans).

#### *3.2.1 Time Evolution*

**Figure 5** presents a sequence of single-pulse fs-pump-probe SAXS patterns recorded upon exposure with a fixed peak laser fluence of $\phi_0$ = 1.1 J cm$^{-2}$ at 15 different delay times $\Delta t$, ranging from −10 ps to +2000 ps. At each delay, SAXS patterns of six identical experimental repetitions were averaged. For background correction of all SAXS patterns, an image averaged over three frames recorded at −10 ps, −2 ps, and 0 ps was subtracted.

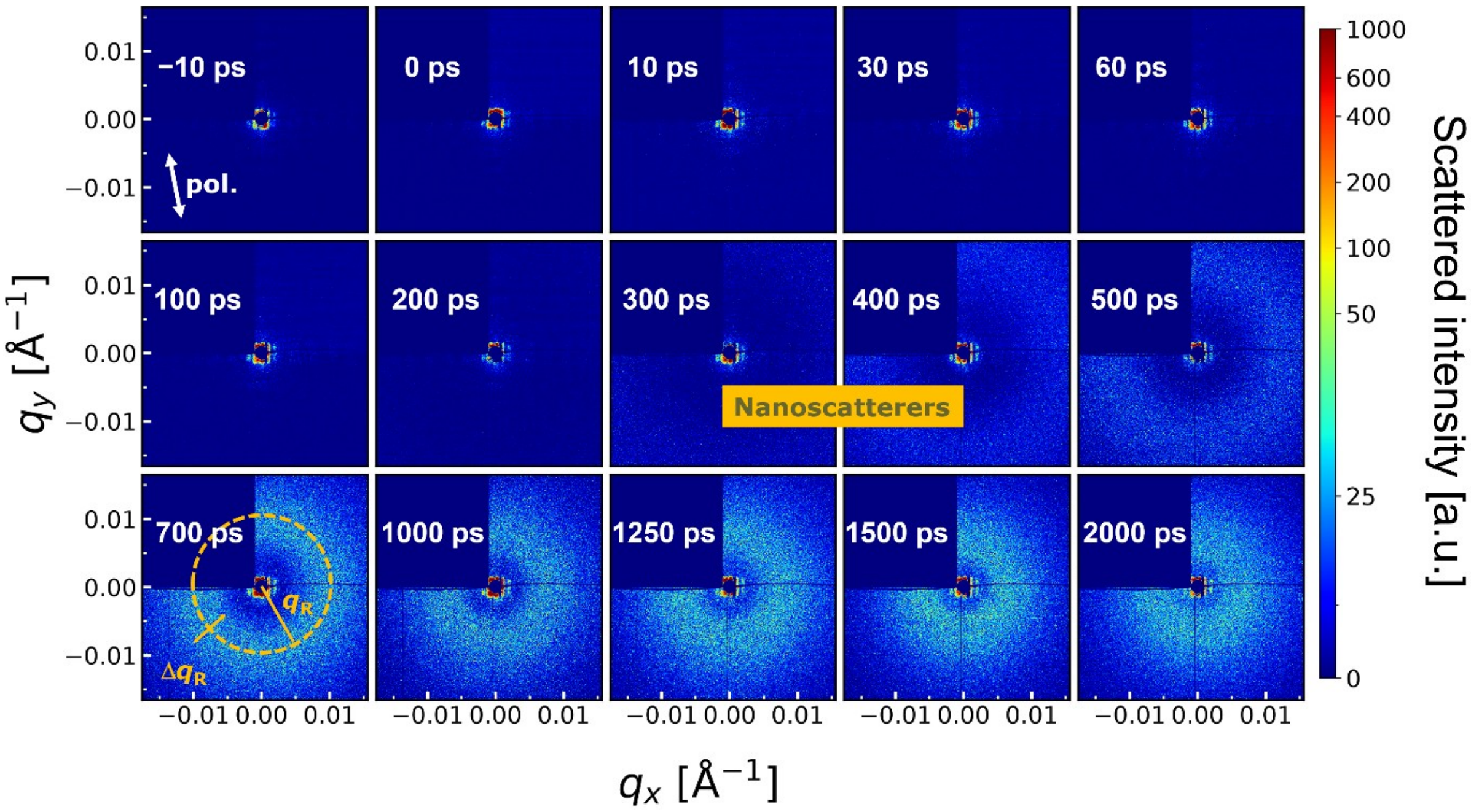

Figure 5: Single-pulse fs-pump-probe SAXS patterns recorded after exposure at a fixed laser fluence of $\phi_0$ = 1.1 J cm$^{-2}$ for 15 different delays $\Delta t$ between −10 ps and +2000 ps. All images share a joint color scale. The white double-arrow in the frame for $\Delta t$ = −10 ps displays the projection of the direction of the linear laser beam polarization onto the sample surface. The ring radius $q_R$ and its FWHM $\Delta q_R$ are exemplary marked in the frame for $\Delta t$ = +700 ps.

For delays $\Delta t \leq$ +200 ps, no relevant scattering signals were recorded. Only a very weak central cross-shaped diffraction signal can be observed along the $q_x$- and $q_y$-axes, arising from diffraction of the outer wing of the spatial X-ray probe beam profile at the edges of the square-shaped sample windows. Starting at a delay around +200 to +300 ps, a broad ring-shaped scattering pattern moves inwards toward smaller $q$-values, while at the same time its intensity increases. It consists of a single ring ("R"). Non-circular transient signatures of LIPSS could not be detected at this moderate laser fluence level.

Such a ring describes a correlation of scattering objects within a certain average distance. Different physical scenarios may lead to the observation of a laser-induced transient ring pattern in SAXS, including (i) the existence of circular (radial) periodic modulations of the mass density (X-ray optical path length variations) across the lateral ($x$-,$y$-)directions, e.g., by the localized excitation of sound waves in the molten $SiO_2$ material (scenario i.a), or by the excitation of surface capillary waves in a laser-molten surface layer (scenario i.b) [26], or via sub-surface nanovoid formation in the $SiO_2$ material (ii). This may occur through the formation of a thin spallative layer that starts to be ejected from the irradiated sample surface, triggering the formation and growth of mesoscopic nanovoids at the inner interfaces upon ablation (scenario ii.a) [38]. Alternatively, the transient ring pattern may arise from the collective X-ray scattering from an ensemble of initially randomly nucleated ($N$ = 1) spherical nanovoids emerging due to tensile stress (cavitation) in the volume of molten silica (scenario ii.b) [15,23]. The two scenarios (ii.a and ii.b) differ primarily in the geometric arrangement and the confinement of the ensemble of nanovoids, their number density and size distribution, as well as in spatially distinct constraints resulting from their different formation mechanisms.

The ring feature reported in [26], which was attributed to the excitation of capillary waves (scenario (i.b)), exhibited a slightly elliptical shape and a weak dependence of the peak position on delay time. Scenarios (ii.a) and (ii.b) both originate from sub-surface thermophysical effects (stress-driven cavitation / evaporation / rarefaction) in the molten material and lead to the formation of "nanovoids"– although the nucleation conditions, shape, arrangement, and material relaxation pathways develop differently. A spallative ablation layer (scenario ii.a) that is caused by a single, sub-ps laser pulse induced rarefaction wave was observed first for bulk metals and semiconductors through fs-time resolved optical microscopy, where the appearance of transient Newton fringes within the ablation region provided direct evidence for an expanding ablation plume with an optically steep front [39-42]. These Newton fringes were initially not observed for dielectric materials [39,43,44] but later found in a narrow laser fluence range above the ablation threshold [45,46]. In our experiments, however, the sub-surface interface between the $SiO_2$ film and the underlying $Si_3N_4$ carrier membrane may lower the film decomposition threshold fluence and spatially confine nanovoid formation at the buried $SiO_2/Si_3N_4$ interface. For further distinguishing between the four scenarios, the characteristics of the ring pattern can be analyzed.

To quantitatively describe the temporal evolution of the ring feature, the recorded SAXS image frames shown in Figure 5 for $\Delta t \geq +300$ ps were azimuthally integrated. Given the perfect circular symmetry, the integration was conducted over the full angular range ($\Delta\varphi = 360°$). Examples of the resulting radial intensity profiles $I_{SAXS}(q_r)$ are shown as black curves in **Figure 6**(a) for three selected delay times $\Delta t =$ +300 ps (top panel), +500 ps (middle panel), and +1500 ps (bottom panel), respectively. The experimental data were decomposed through a least-squares-fit (yellow curve) into a skewed Gaussian peak (orange curve), representing the ring feature, as well as into an exponentially decaying leakage contribution around the beam stop (gray dashed curve).

From the decomposed skewed Gaussian peaks, the corresponding peak position $q_R$ of the ring and its full width at half maximum $\Delta q_R$ (see the middle panel at $\Delta t = +500$ ps) were quantified along with the maximum intensity $I_R$ and the skewness value. Figure 6(b) compiles delay-dependent values of the ring width $\Delta q_R$ (top panel, full red triangles), the peak position $q_R$ (middle panel, full blue squares), and the peak intensity $I_R$ (bottom panel, full black circles), all associated with the left abscissa.

Additionally, values of the *coherent scattering length* $L$ (open green triangles, top panel) and the *correlation distance* $D$ (open violet squares, middle panel) of the nanoscatterers are provided at the right abscissa of Figure 6(b). The correlation distance was calculated as $D = 2\cdot\pi/q_R$ from the ring radius position $q_R$. The coherent scattering length can be derived using a Scherrer-like approach [47] from the radial width $\Delta q_R$ (FWHM) of the ring pattern via $L = 2\cdot\pi\cdot K/[\Delta q_R\cdot\cos(\vartheta)]$, with the X-ray scattering angle $\vartheta$ and a dimensionless shape factor $K$ of the order of unity. Considering the small scattering angles ($\cos(\vartheta) \approx 1$) and setting $K = 1$, the coherent scattering length was approximated as $L \approx 2\cdot\pi/\Delta q_R$.

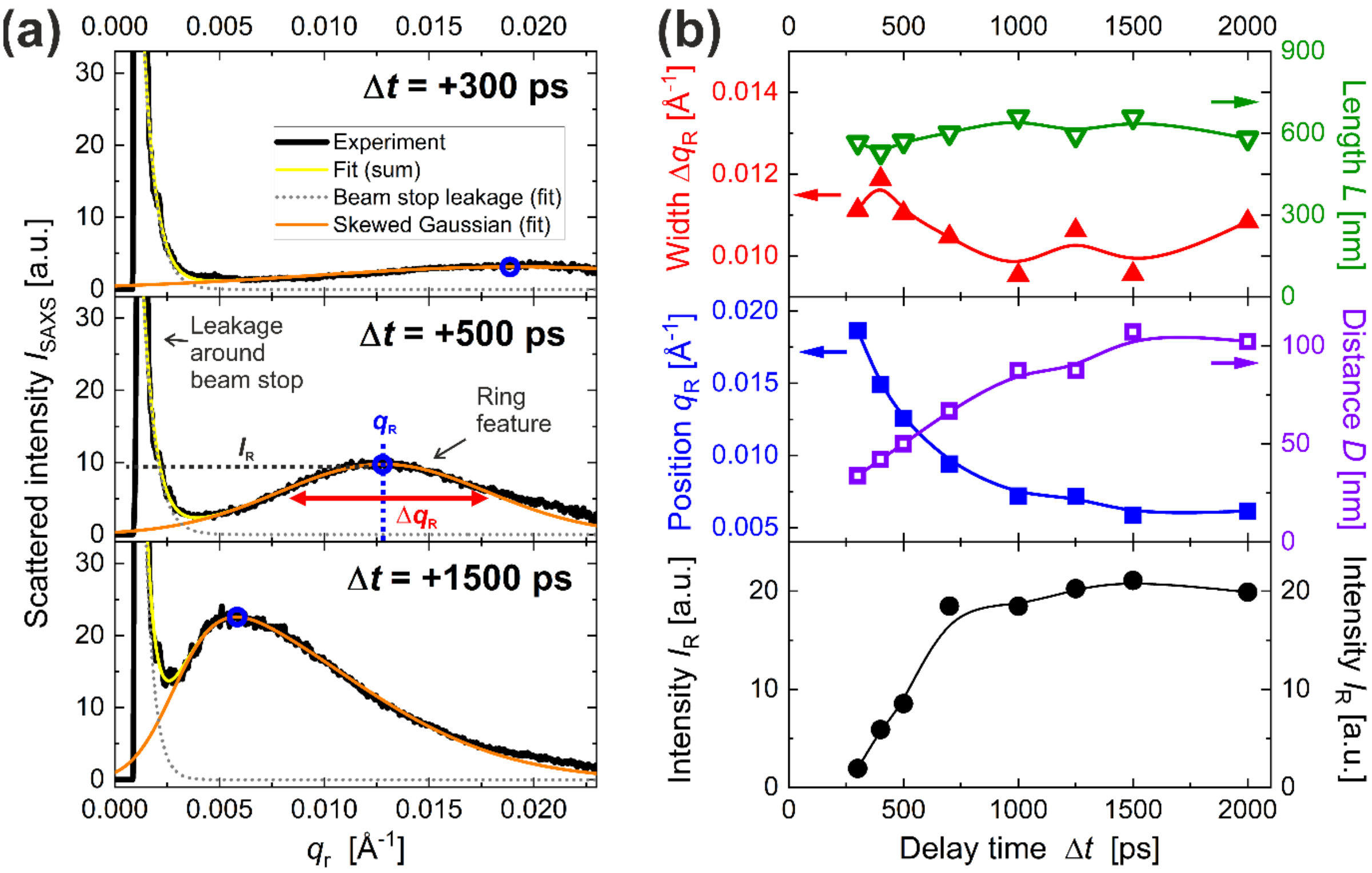

Figure 6: Dynamics of the characteristic ring feature upon single-pulse laser irradiation at a fixed peak fluence $\phi_0$ of 1.1 J cm$^{-2}$. (a) Radial profiles of the scattered intensity $I_{SAXS}$ (black curves) for three selected delay times $\Delta t$ = +300 ps (top panel), +500 ps (middle panel), and +1500 ps (bottom panel), as derived by azimuthal integration over the SAXS patterns presented in Figure 5. Each profile is decomposed by a least-squares fit into a skewed Gaussian peak (orange curve) and an exponentially decaying leakage contribution around the beam stop (gray dashed curve). The blue open circles mark the peak intensity, $I_R$, of the corresponding ring signature. (b) Nanoscatterer characteristics as function of the pump-probe delay time $\Delta t$. Left abscissa: ring width (FWHM) $\Delta q_R$ (top panel), peak position $q_R$ (middle panel), and peak scattering intensity $I_R$ (bottom panel); Right abscissa: coherent scattering length $L$ (top panel) and correlation distance of scattering centers $D$ (middle panel). The lines guide the eye.

The coherent scattering length is essentially constant around $L \approx 600$ nm over the entire delay range. The correlation distance of the nanoscatterers steadily increases linearly from values around $D \approx 30$ nm after $\Delta t = +300$ ps, and saturates around 100 nm for $\Delta t$ exceeding +1.5 ns. The peak intensity $I_R$ of the ring pattern also shows a linear increase with delay time during the first 500 ps, before saturating at a constant value. Therefore, all three quantities do not change after 1.5 ns indicating that the probed material volume has reached a quasi-stationary structural state. Interestingly, the intensity distribution in the $q_r$-direction changes from a rather symmetric Gaussian peak shape (skewness $\gamma = 0.02$) at early delays of $\Delta t = +300$ ps to a more asymmetric Gaussian distribution (skewness $\gamma$ = 3-4) at delay times larger than +1 ns (data curve not shown here for brevity).

Since no asymmetric scattering signatures of LIPSS could be detected in this delay scan performed at the peak laser fluence of $\phi_0 = 1.1$ J cm$^{-2}$, a more systematic analysis was conducted through varying the laser fluence at a fixed pump-probe delay time.

### 3.2.2 *Fluence-dependence*

**Figure 7** provides a series of single-pulse fs-pump-probe SAXS patterns recorded at a fixed delay of $\Delta t$ = +500 ps for ten different laser fluences $\phi_0$ between 0.05 J cm$^{-2}$ and 1.95 J cm$^{-2}$. Such a delay time in the hundreds of ps range allows secondary processes following the ultrafast laser pulse excitation to evolve, e.g., thermal effects, structural transformations or ablation. At each fluence value, SAXS patterns of six identical experimental repetitions were averaged. The averaged image recorded with the laser beam blocked was used for background correction of all SAXS patterns.

At low peak laser fluences between 0.05 J cm$^{-2}$ and 0.45 J cm$^{-2}$, only leakage of the probe beam around the beam stop can be detected in the SAXS patterns, along with a weak, cross-shaped artifact arising from diffraction of the probe beam wings at the edges of the window areas. At laser fluences exceeding 0.66 J cm$^{-2}$, an additional, broad symmetric ring-shaped scattering pattern can be seen, indicating that the single-pulse damage threshold of the $SiO_2$ layer is exceeded. As discussed before, it can be associated with the formation of nanoscattering centers in the silica layer, having a correlation distance of ~48 nm at this delay time. The ring radius and width both depend only weakly on the laser fluence value if the damage threshold is surpassed (a visualization of the variation of the parameters $q_R$, $\Delta q_R$, $I_R$, $L$, and $D$ with the laser fluence, as derived from the ring patterns, can be found in Figure S4 of the *Supporting Information*). Note that at the highest accessible laser fluence of 1.95 J cm$^{-2}$, an additional asymmetric scattering pattern is observed along the laser polarization direction that can be associated with the dynamic formation stage of HSFL-I (see Figure 2), supposedly via the formation of periodic nanocracks in the laser-excited film. The position of highest scattering intensity can be associated with a spatial period around 145 nm. This value deviates by less than 7% from the theoretical prediction of $\lambda/(2 \cdot n_0)$ = 136 nm.

The weak dependence of the ring pattern on the laser fluence $\phi_0$ (also reflected in **Figure S4**) suggests, at first glance, a formation mechanism that does not depend significantly on temperature-sensitive effects, since — given the nonlinear absorption of the ultrashort laser pulse in silicon dioxide — the temperature of the laser-irradiated material changes significantly with fluence. In this context, it is instructive to recall some thermophysical properties of laser-heated amorphous $SiO_2$ glass at temperatures around and above the melting point. At temperatures exceeding the glass transition temperature $T_g$ = 1460 K [48], the silicon dioxide glass softens and melts at $T_m$ = 1986 K [49] before boiling at $T_b$ = 3220 K [49]. In the temperature range between 1600 K and 2400 K, the dynamic viscosity drops by six orders of magnitude from $\eta_{visc} \approx 10^{10}$ Pa·s to $10^4$ Pa·s upon liquifying [50]. In the same temperature range, however, the mass density $\rho$ varies by only 7% from 2.23 g·cm$^{-3}$ to 2.08 g·cm$^{-3}$ [51]. As a consequence, the longitudinal speed of sound $c_{sound} = \sqrt{K_S/\rho}$, with $K_s$ being the coefficient of stiffness (isentropic bulk modulus) of the liquid, varies by less than 2% (6.50 km·s$^{-1}$ (1600 K) to 6.58 km·s$^{-1}$ (2283 K)) in the liquid $SiO_2$ for sound frequencies in the 10 GHz range [52], associated then with sound wavelengths, $\lambda_{sound}$, around 65 nm. The surface tension, $\sigma$, of $SiO_2$ scales approximately linear with the temperature between 0.298 N/m (1773 K) and 0.307 N/m (2073 K) [53].

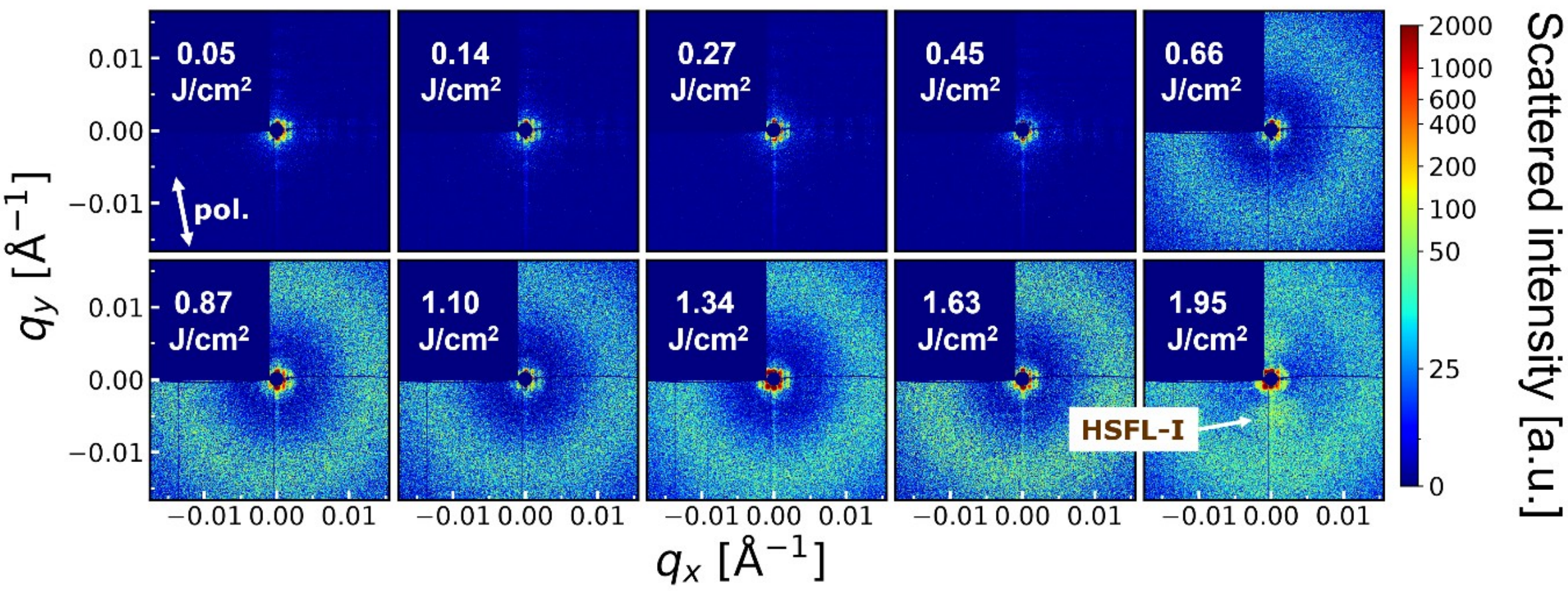

Figure 7: Single-pulse fs-pump-probe SAXS patterns recorded at a fixed delay of $\Delta t$ = +500 ps for 10 different peak laser fluences $\phi_0$ between 0.05 J cm$^{-2}$ and 1.95 J cm$^{-2}$. All images share a joint color scale. The white double-arrow in the frame for $\phi_0$ = 0.05 J cm$^{-2}$ displays the projection of the direction of the linear laser beam polarization onto the sample surface.

The transient characteristics of the ring patterns observed in Figures 5 and 7 are likely not caused by scenarios (ii.a) and (ii.b), outlined above in Sect. 3.2.1, since for both scenarios we would expect a much stronger dependence of the correlation distance $D$ on the laser fluence given the very non-linear absorption behavior of $SiO_2$ at the used pump wavelength of 400 nm. One explanation could be scenario (i.a) with fs-laser-excited longitudinal sound waves as the origin of the transiently scattered ring pattern, as sketched in the following: Above a threshold value of 0.66 J cm$^{-2}$, the high-intensity fs-laser excitation promotes a sufficient number of electrons via nonlinear absorption into the conduction band of $SiO_2$ to allow melting of the film material within a few picoseconds. Already during the laser pulse, a self-ordered modulation of the conduction band electron density manifests due to scattering at the surface roughness, along with near-field interference effects. Once excited, the laser excited electrons start to transfer their energy to the lattice of the solid via electron-phonon relaxation. At the maxima of the electron density distribution, more energy is deposited in the electronic system, in turn resulting in a locally enhanced heating of the material. At these localized sites ("hot spots"), large pressure gradients emerge rapidly that initiate the excitation of longitudinal sound waves. Given the extremely short fs-laser pulse duration, a broad spectrum of sound frequencies is synchronously excited. These sound waves then propagate as spherical waves at the corresponding speed of sound around the hot spots into the surroundings of the softened material. The density modulations associated with these spherical sound waves in the lateral ($x$-, $y$-) directions then manifest as a ring pattern in the SAXS image. The higher sound frequencies (smaller temporal and spatial periods) become visible first at smaller delay times, $\Delta t$. Only a single ring is observed in SAXS because, at all frequencies, density fluctuations related to the sound waves exhibit a sinusoidal modulation to a good approximation. The sound waves are triggered at the hot spots and then propagate into colder regions while being damped. Such longitudinal sound waves are strongly damped in the liquid $SiO_2$ with an attenuation length $\ell_{\text{sound}} \propto \lambda_{\text{sound}}^2$ [54]. The correlation distance $D$ can be associated with the spatial period of the sound waves, and the coherent scattering length, $L$, reflects the lateral size of the probed region from which coherent X-ray scattering emerges. During the emission and propagation of sound waves, heat diffuses, the hot molten silica material cools down, its viscosity increases by several orders of magnitude, and the density modulation state present after ≈ 1.5 ns "freezes in". The changing viscosity and dispersion effects of the longitudinal sound waves may account for the change in the radial intensity distribution of the ring patterns for increasing delay times.

In the second scenario (i.b), capillary waves propagating on the surface of the laser-melted $SiO_2$ layer are the origin of nanoscale X-ray optical path length modulations appearing as transient single-ring patterns in the SAXS. In this scenario, the fs-laser ablation triggers the localized excitation of capillary waves propagating at the surface of the $SiO_2$ layer towards vacuum — again with a broad spectrum of frequencies/periods. Such capillary waves are dispersive, with short wavelength ($\lambda_{cap}$) propagating faster than longer wavelengths (phase velocity $c_{cap} \approx \sqrt{2\pi\sigma/(\rho\lambda_{cap})}$ [55]). Hence, the shorter wavelengths of the excited wave spectrum are detected earlier by the SAXS than the longer ones. Again, the capillary waves are triggered at localized hot spots but then propagate into colder regions while being damped. With increasing delay, the molten silica cools down and largely increases its viscosity. In turn, the damping time of the capillary waves drastically decreases (damping time $\tau_{cap} = \frac{\rho\lambda_{cap}^2}{8\pi^2\eta_{visc}}$ [56]) — eventually freezing the large wavelength surface corrugations at the end of the re-solidification process. This may explain the experimentally decreasing and saturating ring radius along with its changing radial asymmetry (skewness).

Given the similarity between the saturation value of $D$ at 100 nm for large delay times (Figure 6) and the final HSFL-I periods (Figure 3), it is speculated here that the sound wave or capillary wave related density modulations, which are frozen in the film material upon irradiation with a single fs-laser pulse, result in positive inter-pulse feedback for HSFL-I formation. This feedback reinforces the topographic generation of spatial periods around 100 nm for the subsequent laser pulses.

To further distinguish between the four different scenarios ((i.a) laser-induced sound waves, (i.b) capillary surface waves, (ii.a) spallative layer-constrained ablation, and (ii.b) statistically distributed spherical nanovoids), complex, time-dependent, opto-mechanical [57] or opto-hydrodynamical modelling [15,58] or Molecular Dynamics (MD) simulations [38,59], in conjunction with time-resolved pump-probe GISAXS experiments [60,61] (which are sensitive to surface effects) or cross-sectional transmission electron microscopy (TEM) would be required — all, however, beyond the scope of this work having its focus on LIPSS.

### *3.3 Theoretical Analysis of LIPSS on $SiO_2$*

To better understand the experimentally observed SAXS patterns and the formation of LIPSS, two different theoretical approaches were employed. The first approach is based on the first-principles theory of the absorption of electromagnetic radiation at a microscopically rough surface and was developed during the 1980's by J.E. Sipe and co-workers at the University of Toronto (Canada). This analytical method is presented in Sect. 3.3.1 and represents the currently most widely accepted theory of LIPSS. To overcome limitations imposed by certain approximations in this theory, an extended computational analysis based on *Finite-Difference Time-Domain* (FDTD) calculations for solving Maxwell's equations numerically is complemented in Sect. 3.3.2.

#### *3.3.1 Sipe-Drude Model*

The pioneering work of John E. Sipe and co-workers (in the following referred to as "Sipe theory") starts from Maxwell's equations and uses the Green's formalism to derive an integral equation for the dielectric polarization density at a microscopically rough surface [62]. The theory represents a general scattered-field model that includes the possible excitation of SEWs, such as resonantly laser-excited *Surface Plasmon Polaritons* (SPP), and their interference with the incident optical radiation. Sipe's theory can predict possible wave vectors $\boldsymbol{k}$ (spatial frequencies $\boldsymbol{q}$) of LIPSS as a function of laser irradiation parameters (polarization direction, wavelength $\lambda$, angle of incidence $\theta$) and surface parameters (bulk dielectric permittivity $\tilde{\varepsilon}$ and surface roughness parameters). Mathematically, the theory provides an analytical expression for the inhomogeneous deposition of optical energy into the irradiated material (limited to the near-surface layer having a thickness of ~$\lambda$) that can be written as [62]

$$\text{Absorption} \sim \eta(\boldsymbol{k}) \quad (1)$$

Herein, $\eta$ is a scalar response function (the unitless "efficacy factor"), that describes the efficacy with which the surface roughness (represented by $\boldsymbol{q}$) can absorb optical radiation. Peaks in the efficacy factor can occur at specific spatial frequencies of the surface roughness. These peaks can then be used to evaluate the associated spatial periods $\Lambda$ via $\Lambda = 2\pi/|\boldsymbol{q}|$. LIPSS are typically observed where $\eta$ exhibits large amplitudes and/or strong variations, usually associated with its maxima/minima. Sipe's theory provides a coherent explanation for the formation of near-wavelength-sized LIPSS on solids and has become a standard theoretical framework for describing LSFL-I and LSFL-II.

However, some fundamental limitations of Sipe's theory must be mentioned. First, no laser-induced material response is included. This also concerns potential intra-pulse feedback effects, wherein relevant surface/material properties are altered already during the laser irradiation. Second, the model does not consider inter-pulse feedback (changes between individual irradiation events), e.g., through multi-pulse irradiation, altered surface topographies, etc. Third, due to certain mathematical approximations, the absorption of small-scale surface structures with periods of less than ~$\lambda$/5 may not be properly described in the theory. Nevertheless, the efficacy factor theory is currently the most fundamental and widely accepted theory of LIPSS, capable of explaining many experimental observations.

For materials strongly absorbing the laser wavelength (such as semiconductors and metals) Sipe's theory typically predicts a symmetric pair of sickle-shaped features in a two-dimensional map, $\eta(q_x, q_y)$, that is dominating the optical absorption. These sickles represent LSFL-I in reciprocal space, with an orientation perpendicular to the linear laser beam polarization and spatial periods close to the laser wavelength ($\Lambda_{LSFL\text{-}I} \approx \lambda$) in real space. If the irradiated material is additionally plasmonically active at the irradiation wavelength ($\mathrm{Re}\{\tilde{\varepsilon}(\lambda)\} < -1$), the corresponding peaks in $\eta$ are very narrow due to the resonant absorption of the radiation.

On transparent materials that are only weakly absorbing the laser wavelength (dielectrics), another dominant type of LSFL (LSFL-II) is predicted [62] and is often observed experimentally [13]. These surface structures are oriented parallel to the laser beam polarization and feature sub-wavelength spatial periods close to $\Lambda_{LSFL\text{-}II} \approx \lambda/\mathrm{Re}(\sqrt{\tilde{\varepsilon}}) = \lambda/n$ (with $n$ being the real-valued refractive index of the material) in real space. According to Sipe's theory, LSFL-II are related to so-called *Radiation Remnants* (RR), which originate from a specific non-propagating electromagnetic mode located close to the rough surface [63]. This RR mode can redistribute energy from the incident optical radiation to the material at specific spatial frequencies of the surface roughness.

Although the angle of incidence, polarization direction, and surface roughness of the sample can be considered constant during irradiation by a single ultrashort laser pulse, the complex-valued dielectric permittivity $\tilde{\varepsilon}$ may undergo significant changes, particularly when bandgap materials (dielectrics or semiconductors) are exposed to high peak intensities that enable multiphoton absorption processes between the valence and conduction bands of the irradiated solid. In the context of LIPSS, this intra-pulse feedback mechanism was studied in detail for two different materials (ZnO and Si), by combining Sipe's theory with a Drude model [64,65]. This allowed to quantify the change of the dielectric permittivity $\Delta\tilde{\varepsilon}$ as a function of the number density of electrons, $N_e$, promoted into the conduction band of the material upon laser excitation. This extension of Sipe's theory is, therefore, called *Sipe-Drude model* and has been adopted for other materials, including $SiO_2$ [31].

Following that approach, the dielectric permittivity of the laser-excited $SiO_2$ was modeled via $\tilde{\varepsilon}^* = \tilde{\varepsilon} + \Delta\tilde{\varepsilon}_{\mathrm{Drude}}$. The first term $\tilde{\varepsilon} = \epsilon_r + i \cdot \epsilon_i$ is the dielectric permittivity of the non-excited bulk material, while the second additional Drude term is given by [66]

$$\Delta\tilde{\varepsilon}_{\mathrm{Drude}} = \frac{-e^2 N_e}{\epsilon_0 m_{\mathrm{opt}}^* m_e \omega^2 \left[1 + \frac{i}{(\omega\tau_D)}\right]}, \quad (2)$$

with the electron charge, $e$, the vacuum dielectric permittivity, $\epsilon_0$, the free electron mass, $m_e$, the optical effective mass, $m_{\mathrm{opt}}^*$ of the electrons in the conduction band, the angular frequency of the laser radiation, $\omega$, and the Drude damping time, $\tau_D$. Values of $m^*_{opt}$ = 0.49 and $\tau_D$ = 0.4 fs for fs-laser excited fused silica were taken from Ref. [67]. The bulk dielectric permittivity of the non-excited material is calculated from the refractive index, $n$, and the extinction coefficient, $\kappa$, of the material through $\tilde{\varepsilon} = \tilde{n}^2 = (n + i\cdot\kappa)^2$ as $(1.47 + i\cdot 0)^2 \sim 2.16$ at λ = 400 nm [68].

We combined this Drude model implementation with Sipe's theory to study the impact of the electron density in the conduction band of laser-excited (bulk) silica under the conditions of our SAXS experiments and within the range of $N_e$ up to 7 × $10^{22}$ $cm^{-3}$. This range was chosen to cover the entire range from below to above the surface damage threshold. The latter is often associated with the criterion that, at a critical electron density, $N_{cr}$, associated with material damage, the laser angular frequency equals the plasma frequency of the laser-induced free electron plasma in the conduction band of the solid. This leads to resonant absorption of the laser radiation in the electron plasma at $N_{cr} = \epsilon_0 m_e \omega^2/e^2 \approx 7 \times 10^{21}$ $cm^{-3}$ at 400 nm wavelength. For details of the Sipe model implementation the reader is referred to [69,70], software codes are publicly available via GitHub [71].

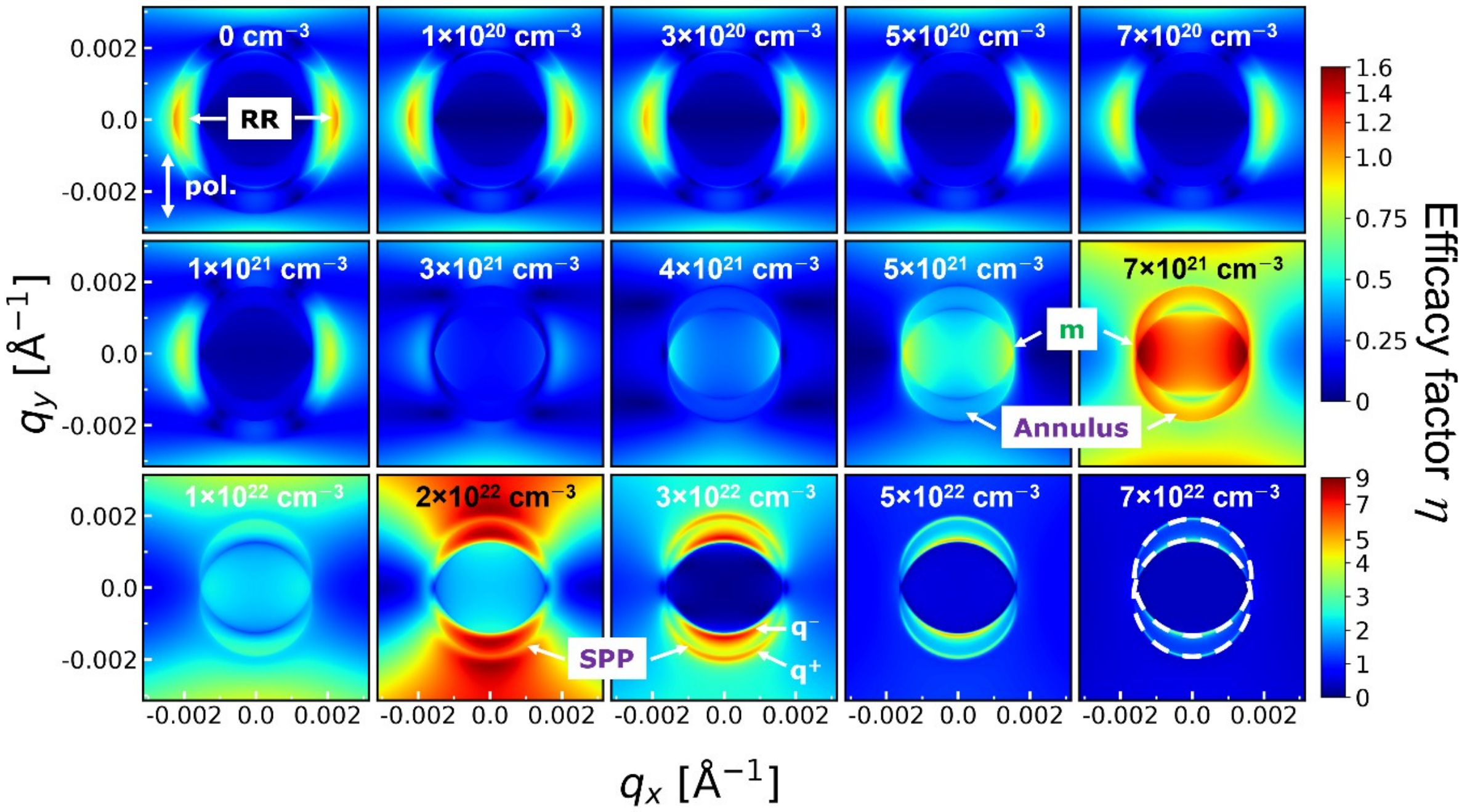

Figure 8: Two-dimensional color maps of the efficacy factor η (Sipe-Drude model) for bulk silica as a function of LIPSS spatial frequencies ($q_x$, $q_y$) at different excitation levels, $N_e$, of the material [$\theta$ = 11.5°, p-polarization, λ = 400 nm]. For improved visibility, the values of $\eta$ are encoded in a joint color scale for the upper two image rows [$0 \leq \eta \leq 1.6$] and for the lowest row [$0 \leq \eta \leq 9$], respectively. The projection of the linear laser beam polarization direction onto the sample surface is indicated in the map for $N_e$ = 0 $cm^{-3}$. The two intersecting white dashed circles in the map for $N_e$ = 7 × $10^{22}$ $cm^{-3}$ mark the positions permitted by momentum conservation in a simple SEW model, ruled by the angle of incidence of the laser radiation. Abbreviations: RR – Radiation Remnants; m – mixed features; SPP – Surface Plasmon Polaritons.

**Figure 8** assembles a series of two-dimensional efficacy factor maps, $\eta(q_x, q_y)$, calculated by the Sipe-Drude model for varying degrees of laser-induced material excitation, represented by 15 different electron densities, $N_e$, in the conduction band of bulk $SiO_2$, ranging from 0 to 7 × $10^{22}$ $cm^{-3}$. The maps were computed for the wavelength λ = 400 nm, an angle of incidence $\theta$ = 11.5°, a shape factor $s$ = 0.4 and a filling factor $f = 0.1$. The latter are the standard roughness parameters in Sipe's theory, representing spherically shaped islands as scattering centers [62].

For the non-excited material ($N_e = 0$ cm$^{-3}$), the $\eta$-map is dominated by a pair of sickle-shaped, double-arc features with coinciding maxima centered around ($q_x \pm 2\cdot\pi\cdot n_0/\lambda = \pm 2\cdot\pi\cdot 1.47/(4000$ Å$) \approx 0.0023$ Å$^{-1}$, $q_y = 0$) in reciprocal space, as indicated by the arrows. These features originate from the excitation of Radiation Remnants (RR, see above) and correspond to LSFL-II, with lateral periods around $\Lambda \sim \lambda/n_0 = 400$ nm/1.47 ~272 nm and an orientation parallel to the laser beam polarization in the spatial domain. For increasing electron densities between $1 \times 10^{20}$ cm$^{-3}$ and $3 \times 10^{21}$ cm$^{-3}$, the position of the maximum of RR-features shifts toward smaller spatial frequencies, $q_x$ (associated with larger spatial periods), while its peak amplitude decreases, and the contour starts to blur and widen. The shift of the peak position can be qualitatively explained by the transient change of the optical refractive index, $n$, which, according to the Drude model, drops in the aforementioned electron density regime for increasing $N_e$ from the value $n_0$ of the non-excited material (data not shown here).

For $N_e$ exceeding $3\times10^{21}$ cm$^{-3}$, the RR-features vanish and transform into another, orthogonally arranged set of double-arc peaks at even higher electron densities ($N_e > 1\times10^{22}$ cm$^{-3}$) when the optical properties of the material turn metallic due to the large number of laser-excited conduction band electrons. These peaks can be associated with the excitation of SPP, when $\mathrm{Re}\{\tilde{\varepsilon}\}$ turns smaller than −1 [72], and represent the classical LSFL-I signatures for non-normal incident radiation. This constraint of SPP-activity is fulfilled for $N_e > N_{e,SPP} = 1.383 \times 10^{22}$ cm$^{-3}$, here.

In a narrow transitional carrier density range between ~$5 \times 10^{21}$ cm$^{-3}$ and ~$1 \times 10^{22}$ cm$^{-3}$ another set of dominant features appears, with maxima centered around $q_x \approx \pm 0.0016$ Å$^{-1}$, $q_y = 0$, surrounded by a broad annulus, see the maps calculated for $N_e = 5 \times 10^{21}$ cm$^{-3}$ and $7 \times 10^{21}$ cm$^{-3}$. These two maxima can be associated with spatial periods close to the laser irradiation wavelength and represent periodic patterns oriented parallel to the laser beam polarization. These characteristic peaks were predicted by Déziel et al. [73] as "m-features" (m – mixed) and manifest when $\mathrm{Re}\{\tilde{n}\}$ equals $\mathrm{Im}\{\tilde{n}\}$ ($n = \kappa$) if the ratio $\gamma/\omega$ of the Drude collision frequency ($\gamma = 1/\tau_D$) and the laser angular frequency ($\omega$) is small. For our parameters, this equality occurs at electron densities of $N_e \approx 9.5 \times 10^{21}$ cm$^{-3}$. The m-features vanish when the laser-excited material properties are driven further towards SPP-activity at larger electron densities.

The absolute efficacy factor amplitude (absorption) for the resonant SPP-related sample excitation channel is largest in the map calculated for $N_e = 2 \times 10^{22}$ cm$^{-3}$ and decreases again at higher electron densities. The double-arcs represent the $q^+$-/$q^-$-modes already observed in the SAXS patterns in Figure 2. Their $q_y$-positions at $q_x = 0$ can be associated with $\Lambda^{\pm} = \lambda/[\lambda/\lambda_{SPP} \pm \sin(\theta)]$, with $\lambda_{SPP}$ being the spatial period of the SPP, as predicted by the simple SEW scattering model [26,74,75]. For our case of a laser-excited sample-vacuum interface, the ratio of the two wavelengths is given by $\lambda/\lambda_{SPP} = \mathrm{Re}\{[\tilde{\varepsilon}/(\tilde{\varepsilon}+1)]^{1/2}\}$ [76]. For SPP active metals and strongly excited semiconductors/dielectrics it can be approximated by $\lambda/\lambda_{SPP} \approx 1$. The expression for the $q^+$-/$q^-$-modes then accounts to $\Lambda^{\pm} = \lambda/[1 \pm \sin(\theta)]$ and coincides with the simple surface scattering model [34]. The SEW model uncovers that, due to momentum conservation, the permitted wave vectors of the surface-scattered electromagnetic waves (LSFL-I) are determined by the angle of incidence and must be located on two intersecting circles, as indicated in the figure by white dashed lines in the $\eta$-map for $N_e = 7 \times 10^{22}$ cm$^{-3}$. From the efficacy factor maps it becomes clear that its anatomy and electron density dependent morphogenesis of the identified features deserve further quantitative analysis.

To compare the evolution of the characteristic peaks as a function of electron density, the positions of their maximum values (resulting in the predicted LIPSS periods) and the corresponding amplitudes, $\eta_{max}$ (as a measure for the strength of optical absorption), were quantified for the RR-features (LSFL-II), the m-features, and the SPP-features (LSFL-I). The results are displayed in **Figure 9** as a function of laser-excited electron density, $N_e$, in a semi-logarithmic representation. Panel (a) presents the curves for the predicted spatial periods, Λ, while panel (b) shows the corresponding curves of $\eta_{max}$. To account for variations in the periods of the different features, we have empirically chosen a 95% of maximum criterion to evaluate their widths, and have quantified the corresponding minimum and maximum spatial periods accordingly. The corresponding period ranges are marked as color-shaded bands surrounding the data points associated with the peak positions.

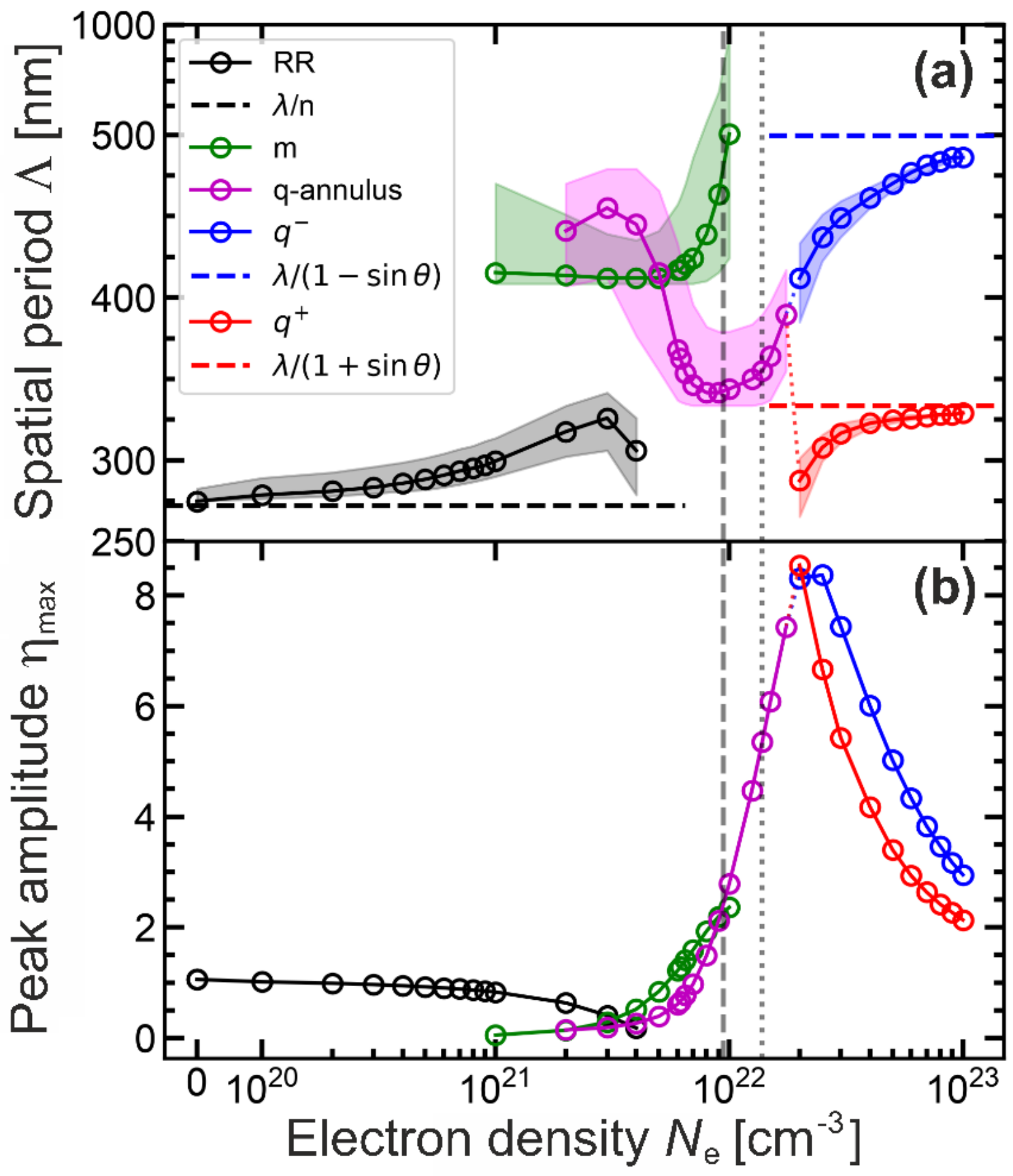

Figure 9: Spatial periods (a) obtained from the positions of the different features (RR, m, q) in the $\eta$-maps (peak values as data points and upper and lower limits obtained from a 95% criterion as color-shaded bands) and the associated maximum values of the efficacy factor (b) for the different features as a function of the excitation level, $N_e$, of the laser-excited silica. The horizontal dashed black line in (a) marks the value $\lambda/n_0$ associated with LSFL-II, while the blue and red lines delineate the limits predicted by the SEW model for the supra- ($q^-$) and sub-wavelength ($q^+$) LSFL-I modes. The vertical dashed and dotted lines indicate the criteria for $N_e$ establishing $n = \kappa$ and $N_e = N_{e,SPP}$, respectively. Note the logarithmic electron density scale. The lines guide the eye.

For the RR features, the period increases from the value for the non-excited material ($N_e = 0$ cm$^{-3}$) of $\lambda/n_0 = 272$ nm (indicated by a horizontal dashed black line in Figure 9(a)) to values up to ≈ 330 nm at $N_e \approx 4 \times 10^{21}$ cm$^{-3}$, while the amplitude of $\eta_{max}$ steadily drops from a value around one toward zero (in Figure 9(b)). At electron densities between $2 \times 10^{21}$ cm$^{-3}$ and $1 \times 10^{22}$ cm$^{-3}$, the m-features can be identified featuring steeply increasing spatial periods between 420 and 500 nm along with $\eta_{max}$ values rising up to 2.36 when $n(N_e)$ equals $\kappa(N_e)$ at $N_e \approx 9.5 \times 10^{21}$ cm$^{-3}$. In a similar electron density range, the characteristic annular-shaped feature emerges, showing a similar increase in the $\eta_{max}$ amplitudes. However, the periods associated with the ring maxima exhibit a very different behavior when dropping from a maximum period of 455 nm at $3 \times 10^{21}$ cm$^{-3}$ through a minimum of 342 nm at around $8.5 \times 10^{21}$ cm$^{-3}$ before increasing again. This rising part of the curve then splits into two branches for the $q^+$- and $q^-$-modes of the LSFL-I features once the material becomes SPP-active at electron densities exceeding $N_{e,SPP} = 1.38 \times 10^{22}$ cm$^{-3}$ (see the vertical dotted line in Figure 9(a) and the discussion above). The periods of the $q^-$-branch steadily continue the trend of the ring feature, while those of the $q^+$-branch show a discontinuous drop close to the threshold density for SPP excitation. At even larger electron densities, both branches steadily increase with saturation towards the limiting values predicted by the SEW model, i.e., $\Lambda^+ = \lambda/[1 + \sin(\theta)] \approx 334$ nm and $\Lambda^- = \lambda/[1 - \sin(\theta)] \approx 500$ nm, respectively (both indicated by dashed blue and red horizontal lines, respectively). At electron densities close to $N_{e,SPP}$, the peak efficacy factor $\eta_{max}$ exhibits very large amplitudes between 8 and 9 for both branches, indicating the resonant energy absorption in the laser-excited material through the SPP excitation channel. The maximum amplitudes of both LSFL-I modes decrease again to values below 3 as $N_e$ exceeds $10^{23}$ cm$^{-3}$. Except for $N_e$ values close to $N_{e,SPP}$, the $q^-$-mode shows somewhat larger efficacy factor values than the $q^+$-mode at such high degrees of laser excitation (Figure 9(b)).

In general, the Sipe-Drude model, which concerns bulk materials and uses a global dielectric permittivity value, can explain several of our experimental SAXS observations, e.g., the presence of LSFL-II and LSFL-I signatures, including the two $q^+$-/$q^-$-modes, that can be associated with RR and the excitation of SPPs, respectively. With the help of the electron densities, these features can be related to optical properties of the laser-excited material, indicating the dominance of the RR excitation channel in the transparency regime of the sample. Efficient SPP excitation emerges in a specific electron density range when the material turns optically metallic.

Specifically, the SPP-driven LSFL-I dominate the LIPSS formation if the laser-induced electron density is globally driven beyond $N_{e,SPP}$, either via intra-pulse nonlinear absorption at sufficiently high laser fluences or via inter-pulse incubation effects that increase the optical absorption by generating electronic defect states upon multi-pulse irradiation at moderate to low fluences. LSFL-II may emerge at low fluences and with a large number of laser pulses. This allows incubation to locally increase the absorption coefficient in sub-surface laser interference patterns, eventually leading to the spallation of the overlying film material. Our Sipe-Drude model cannot anticipate the formation of HSFL-I since these small-scale structures are beyond the validity range of the analytical Sipe theory. Therefore, a less constrained theoretical approach is required to properly model these sub-diffraction limit surface structures as well.

### *3.3.2 Finite-Difference Time-Domain (FDTD) Calculations*

Such a generalized approach may be provided through numerical simulations. In the context of fs-laser-induced self-ordered nanostructures in fused silica, in-depth FDTD calculations have been reported by Buschlinger et al. for volume nanogratings [77] and by Rudenko et al. and Déziel et al. to clarify the physical origin of the HSFL-I [15,32]. These studies revealed that HSFL-I originate from pulse-by-pulse incubation-affected collective optical near-field scattering effects of laser excited near-surface defects [15]. This became possible since these FDTD simulations include also local changes of the optical material properties, e.g., through defect-mediated laser-induced plasmonic nanoscale scattering centers excited at the sample surface. Moreover, the sub-wavelength feature size around tens to hundreds of nanometers fall within the validity range of the FDTD model implementation. Hence, the FDTD approach can complement and extend the results of the Sipe-Drude model presented in Sect. 3.3.1.

The FDTD approach, as implemented by J.-L. Déziel [32,73], was used to calculate the self-reconfigured nanoscale fs-laser-induced conduction band electron plasma patterns in fused silica for assessing their real space and reciprocal space characteristics closely below the laser-irradiated sample surface. This three-dimensional FDTD approach numerically solves Maxwell's equations and includes all relevant intra-pulse electromagnetic effects, e.g., linear and nonlinear absorption, tunnel and collisional ionization, near- and far-field scattering effects, interference, plasma optical response, etc. For details of the FDTD calculations the reader is referred to [32], the Python-based software package is publicly available at the *pyplasma* project webpage and via GitHub [78].

As the *pyplasma* FDTD implementation relies on normal incident laser radiation, $\theta = 0°$ was chosen, along with $\tau$ = 50 fs, $\lambda$ = 400 nm, and $\tilde{n}(400 \text{ nm}) = n_0 + i \cdot \kappa_0 = 1.47 + i \cdot 0$ for the non-excited silica material, assuming a band gap energy of 9 eV. The surface roughness of the bulk silica sample was implemented from fractal noise by an additional surface layer with a maximum height ($z$) modulation amplitude of 9 nm and a lateral feature size of up to 100 nm.

A space domain ($x, y, z$) of 6 × 6 × 0.45 µm$^3$ was chosen for the simulations, along with perfectly matched layer (PML) boundary conditions. The spatial discretization steps were 11.7 nm in the lateral $x$- and $y$-directions and 9.0 nm in the beam propagation direction, $z$. A time discretization step of 18.3 as was selected and used for numerical integration over a total time window of 100 fs to provide the electron density pattern $N_e(x, y, z, t = 100 \text{ fs})$, right after the fs-laser pulse interaction with the silica sample. The electron density distribution was calculated for single-pulse irradiation at two distinct laser fluences, mimicking laser excitation conditions that lead to maximum values below or above the critical electron density, $N_{cr}$, respectively.

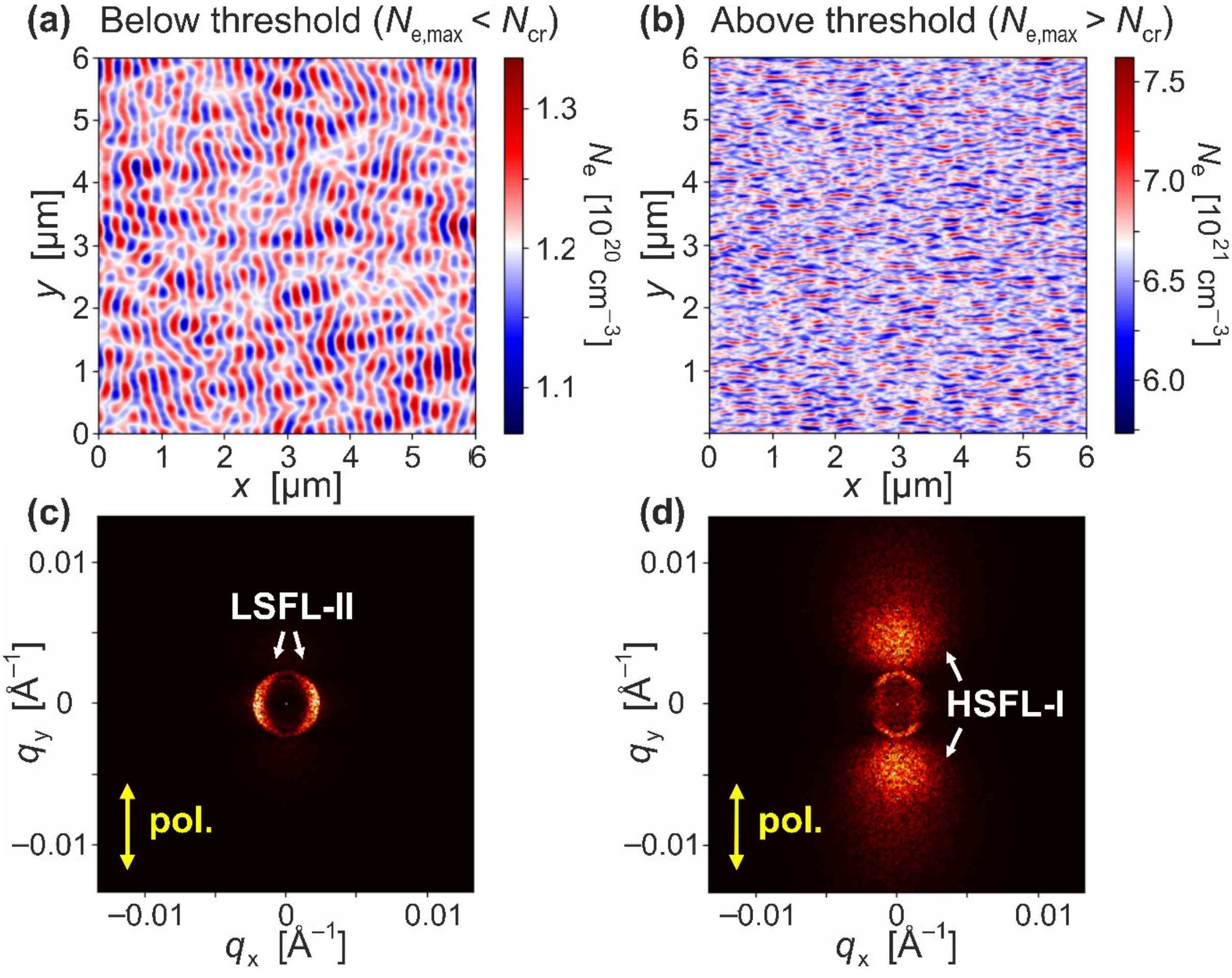

Figure 10: Variations in electron plasma density, $N_e(x, y)$, at a fixed depth of $z$ = 50 nm in the silica sample after interaction with a single laser pulse ($\lambda$ = 400 nm, $\tau$ = 50 fs, $\theta = 0°$) with a fluence of 0.7 J cm$^{-2}$ (a) and 1.6 J cm$^{-2}$ (b). The bottom panels (c, d) provide 2D-FFTs of the plasma patterns displayed above. The direction of the linear laser beam polarization is indicated by yellow double-arrow in (c) and (d).

**Figure 10** visualizes electron density data after the fs-laser pulse excitation at a fluence of 0.7 J $cm^{-2}$ (left column) and 1.6 J $cm^{-2}$ (right column), evaluated 50 nm below the surface plane. The top row panels represent false-color maps of the electron density $N_e(x, y, z = 50$ nm, $t = 100$ fs), and the bottom row panels provide the corresponding 2D-FFTs. Periodic patterns of varying sizes and orientations can be seen. For sub-threshold laser excitation at 0.7 J $cm^{-2}$ (a), the electron density features a pattern with LSFL-II signatures that are parallel to the laser beam polarization. The corresponding spatial periods obtained from the 2D-FFT (c) are $\Lambda \approx 295$ nm, close to the expected value of $\lambda/n_0 = 272$ nm for the RR.

At a fluence of 1.6 J $cm^{-2}$, the maxima of the electron density pattern locally exceed the critical threshold value of $N_{cr}$, $\approx 7 \times 10^{21}$ $cm^{-3}$, while the minima stay below. The electron plasma distribution is dominated by the HSFL-I signature, with a significantly smaller spatial period and oriented perpendicular to the polarization (b). Spatial periods around $\Lambda \approx 150$ nm, close to $\lambda/(2n_0) = 136$ nm, are obtained here from the Fourier transform, represented by the wide, cloud-like lobes in (d). The natural explanation for this periodicity is the formation of standing optical waves along the $y$-axis through the local excitation of nanoplasmas [79]: Light scattered toward the $+y$- and $-y$-directions can interfere, forming a standing wave with nodes that are spatially separated by half the laser wavelength in the surrounding medium ($\lambda/(2 \cdot n_0)$) at every maximum/minimum of the electron plasma density. The 2D-FFT additionally reveals the coexistence of signatures of LSFL, represented by two sharp, sickle-shaped arcs arranged in the same vertical direction. In contrast to the Sipe-Drude efficacy factor calculation presented above, only one pair of arcs is present here, which is fully in line with the SEW model and Sipe's theory for normal laser incidence. The spatial frequencies associated with these arcs represent periods of $\Lambda \approx 300$ nm close to the laser wavelength, $\lambda$, within the silica material ($\lambda/n_0$). This indicates the presence of nanoplasmonic scattering centers within the silica material. Supposedly, their emitted radiation can interfere constructively within the surrounding glass matrix when co-propagating in the direction of the laser polarization, while the coherently laser-driven scattering centers have an average distance close to $\lambda/n_0$. However, the orientation and the period of the associated structures may also be consistent with the q-annulus feature observed in the Sipe-Drude calculations in the transition range to the $q^+$-mode (Sect. 3.3.1), again reflecting the semitransparent to metallic state of the irradiated material.

It was pointed out already by Déziel et al. that the standing-wave explanation for the HSFL-I periodicity is only valid for single- or few-pulse laser irradiations [32]. For multiple-pulse irradiations, inter-pulse feedback additionally comes into play and can explain the large extent to which HSFL-I signatures are observed in the polarization ($y$-)direction in the pulse-by-pulse SAXS experiments (Figure 2). For single-pulse laser irradiation at fluences above the damage threshold, however, the aspect ratios of the periodic nanocracks/HSFL-I features in the direction of the laser polarization and perpendicular to it, are rather similar and close to one, compare Figure 7 (1.95 J $cm^{-2}$) and Figure 10(d).

Although the FDTD numerical simulations were performed for normal incident laser radiation, they successfully resemble the HSFL-I and LSFL-II characteristics ruled by intra-pulse scattering and interference effects in the conduction band electron density. Via energy relaxation processes from the conduction band electrons to the lattice and via multi-pulse irradiation, the HSFL and LSFL then manifest also in the surface topography of the sample, e.g., via melting, void formation, cracking, and incubation-affected ablation. The FDTD approach complements and extends the Sipe-Drude modelling and confirms that electromagnetic effects seed the characteristics and early-stage formation of LIPSS on silica.

### *3.4 LIPSS Formation Stages and Mechanisms*

Summarizing our ultrafast SAXS experiments, we propose the following scenario of fs-LIPSS formation on $SiO_2$ thin films at peak laser fluences below the single-pulse damage threshold. The process can be described in five stages, which emerge with an increasing number of laser pulses.

**Stage I (Incubation phase):**

For laser intensities/fluences moderately below the single-pulse damage threshold and a low number of laser pulses, $N$, electronic defect accumulation occurs in the $SiO_2$ film, which collectively lowers pulse-by-pulse the optical refractive index of the irradiated material. The highest defect density is created at the surface and in the center of the Gaussian laser beam profile where the local intensity is largest. At sufficiently high local laser intensities, these defects may transform into nanometric plasmonic scattering centers driven by the laser pulse. Transient nanoscatterer formation may occur within a few nanoseconds after the laser pulse at sample sites where the locally absorbed optical energy exceeds a certain threshold value (Figure 5).

**Stage II (Seeding of HSFL-I via coherent plasmonic near-field scattering and nanocrack formation):**

Collective near-field scattering of plasmonic nanoscatterers and optical interference leads to a laterally modulated nanoscale intensity pattern (period $\Lambda_{HSFL} \approx \lambda/(2\cdot n)$) that carries the HSFL-I signatures with an orientation perpendicular to the laser beam polarization close to the air/silica interface. This optical interference pattern decays into the film on a depth scale of the order of 100 nm [15,32]. If the locally absorbed laser intensity at the maxima of the interference pattern exceeds a threshold value, localized periodic nanocracks develop within several hundred picoseconds (Figure 7) at the film surface, permanently modifying the sample topography. The near-field scattering effects described above for a single laser pulse, in combination with multi-pulse incubation, i.e., electronic defect accumulation at the interference maxima, trigger an inter-pulse feedback process. This process leads to spatially coherent ordering of the nanocracks at periods close to $\Lambda_{HSFL}$ over the entire center of the irradiated laser spot, while further enhancing the crack depths into the bulk of the silica film. Due to the large local change in the refractive index for X-rays, these topographic changes become visible in the SAXS patterns. As the number and depth of the periodically arranged nanocracks increases, the HSFL-I signature becomes stronger with an increasing number of pulses per spot, while their most frequent period stabilizes around $\Lambda \approx 100$ nm (Figure 2, $N \approx 80$ to 150).

**Stage III (Global plasmonic activity driving LSFL-I surface grating formation):**

When the number of electronic defects accumulated at the surface of the silica film through additional laser pulses reaches a critical density, subsequent laser pulses interact with a surface of widely homogeneous dielectric permittivity featuring transiently metallic properties. If $\mathrm{Re}\{\tilde{\epsilon}\} < -1$ ($\Rightarrow N_e > 1.383 \times 10^{22}$ $cm^{-3}$), SPP can be excited across the air/silica film interface. This results in interference between the SPP electromagnetic fields and the incident laser pulse. This additionally periodically modulates the pattern of absorbed optical energy by the signatures of LSFL-I at spatial periods close to $\lambda$ and with an orientation perpendicular to the laser beam polarization. Due to the non-normal laser-incidence, two different modes ($q^+$ / $q^-$) can manifest with spatial periods around $\Lambda^{\pm} = \lambda/[1 \pm \sin(\theta)]$ (Figures 8 and 9). Through energy transfer to the lattice of the material and localized ablation, the LSFL-I topographic surface grating becomes visible in the SAXS patterns at spatial frequencies of the $q^+$-mode close to the beam stop (Figure 2, $N > 50$). The global plasmonic activity may be reached even for single laser pulse irradiation at sufficiently high intensities/fluences, allowing the necessary laser-induced electron density to be exceeded at times close to the temporal pulse maximum.

**Stage IV (Sub-surface LSFL-II seeding and HSFL/LSFL removal through film ablation):**

The laser radiation scattered by surface defects also propagates into the film material and can interfere with the incident laser pulse propagating through the (semi-)transparent film material. At sub-surface depths around 50 nm (Figure 10), a rotated interference pattern emerges that carries the spatial signature of LSFL-II, i.e., spatial periods close to $\Lambda_{LSFL\text{-}II} \approx \lambda/(2\cdot n)$ and with an orientation parallel to the laser beam polarization. This sub-surface seeding mechanisms of LSFL-II acts on bulk dielectric materials [15,31] and also manifests on thin transparent films [33]. Given the negligible associated mass density changes via incubating electronic defects, this pattern remains undetectable by the probing X-ray wavelength and is inaccessible through SAXS. However, again, pulse-by-pulse electronic defect accumulation enables additional sub-surface absorption of optical radiation. This facilitates the ablation of the covering film material with near surface HSFL-I and LSFL-I, leaving behind an LSFL-II corrugated surface. After ablation, the signature of this surface topography becomes visible in the SAXS patterns of Figure. 2 ($N \approx 150$).

**Stage V (Film removal to steady state conditions):**

If the laser fluence is high enough, the silica film may be removed from the underlying carrier membrane with additional laser pulses. Due to the spatially Gaussian laser beam profile, only HSFL-I, representing the surface morphology formed at the lowest laser fluences, "survive" in the low-fluence regions of the laser-spot. They can then be probed with the wings of the X-ray pulses (Figure 4), remaining as the dominant feature in the SAXS pattern (Figure. 2, $N > 200$). See also the "movie" S3 related to the pulse number series of Figure 2 in the *Supporting Information* for following the disappearance of the LSFL signatures at such large numbers of laser pulses.

## 4. Conclusions

Pump-probe fs-SAXS experiments in transmission geometry (probe beam: 0.83 nm, 25 fs, 0°, 40 µm) enabled the study of the formation of transient and permanent sub-micrometric nanostructures (nanoscatters, nanocracks, LIPSS) in membrane-supported 200 nm thick $SiO_2$ films with sub-ps temporal resolution. The setup also provided the capability to track *in-situ* the pulse-by-pulse evolution of the LIPSS (HSFL-I. LSFL-I, LSFL-II) at a fixed irradiation site. Using this approach, we demonstrate for the first time, through multi-pulse fs laser irradiation (400 nm, < 50 fs, 11.5°) at a fixed peak laser fluence of 0.27 J $cm^{-2}$ that, during the first 100 laser pulses, positive inter-pulse feedback effects lead to significant ordering of the HSFL-I topographies with spatial periods between 50 nm and 100 mn, only. Upon single laser pulse irradiation (pump beam: 400 nm, < 50 fs, 11.5°, 90 µm, p-pol.) at laser fluences exceeding a threshold value (≈ 0.66 J $cm^{-2}$), nanoscatterers with correlation distances between 30 and 100 nanometers formed within the silica film after delay times of 300 ps to 2 ns, respectively. Four physical scenarios were outlined as potential origin of the optically excited nanoscattering centers, including longitudinal sound or capillary surface waves, spallative layer-constrained ablation, or randomly distributed spherical nanovoids in the bulk of the laser-induced molten silica film. At a larger laser fluence of around 2 J $cm^{-2}$, periodic nanocracks were detected 500 ps after the laser pulse. These nanocracks already carry the spatial signature of HSFL-I with periods of approximately 150 nm, formed upon multi-pulse irradiation through collective coherent near-field scattering and interference of transiently laser-excited nanoplasmonic defect centers at the silica/air interface. Our time-resolved study identified the nanoscattering centers with a distinct formation threshold fluence as a transient precursor to HSFL-I on dielectrics and demonstrated, for the first time, that this type of sub-diffraction limited LIPSS can form in less than one nanosecond. With the help of two complementary theoretical approaches through the analytical Sipe-Drude model and numerical FDTD simulations, the features observed in fs-SAXS could be related to physical processes. The associated laser-induced carrier density ranges were identified, and the relevance for incubation effects upon multi-pulse irradiation at low to moderate laser fluences was proven.

## 5. Materials and Methods

### *5.1 Sample Preparation and Characterization*

The $SiO_2$ layers were deposited at the German Federal Institute for Materials Research and Testing (BAM, Berlin, Germany) using an electron beam evaporation physical vapor deposition (PVD) process under high vacuum conditions (total pressure $3\times10^{-5}$ mbar). The coating system employed was a cluster coating system, model CS 730 ECS, manufactured by von Ardenne Anlagentechnik GmbH (Dresden, Germany). The substrates (sample carriers) consisted of commercially available 300 × 300 µm$^2$ sized amorphous $Si_3N_4$ membrane windows (thickness 300 nm) arranged in large arrays and supported by a silicon wafer frame (Silson Ltd., Southam, England). This design allows for easy replacement of the sample in case of damage caused by laser or XFEL radiation. The evaporation material consisted of $SiO_2$ granules with a purity of 99.99%. Evaporation was carried out from a copper crucible at an electron acceleration voltage of 6 kV and a beam current of ≈ 22 mA, resulting in a deposition rate of 10 Å/s. To achieve a uniform thickness distribution, the sample holder with the substrates was moved during the coating process. The time required for deposition of a ≈ 200 nm thick amorphous $SiO_2$ layer on the entire sample carrier was determined from previous thickness calibration experiments using test samples along with a tactile profilometer (Bruker Dektak XT, Billerica, USA). The chosen thicknesses of the $SiO_2$ layer and $Si_3N_4$ membrane ensured sufficient transmission for the selected X-ray photon energy (1.5 keV; see below).

The surface topography of the 200 nm thick $SiO_2$ layers deposited on the $Si_3N_4$ membranes was characterized with a scanning white light interference microscope (SWLIM) NexView from Zygo/AMETEK corporation (Middlefield, USA), allowing to acquire 3D images with a vertical (*z*-) resolution in the range of approximately 1 - 2 nm. With the chosen 20× Mirau interferometer objective, a lateral optical resolution of 0.7 µm was realized within a measured region of 420 × 420 µm² (camera pixel resolution 0.4 µm). Cross-sectional line scans using the OEM software Mx (Zygo) confirmed a smooth $SiO_2$ layer morphology and layer/membrane deformation of less than 5 nm over the entire window size, see **Figure 11**.

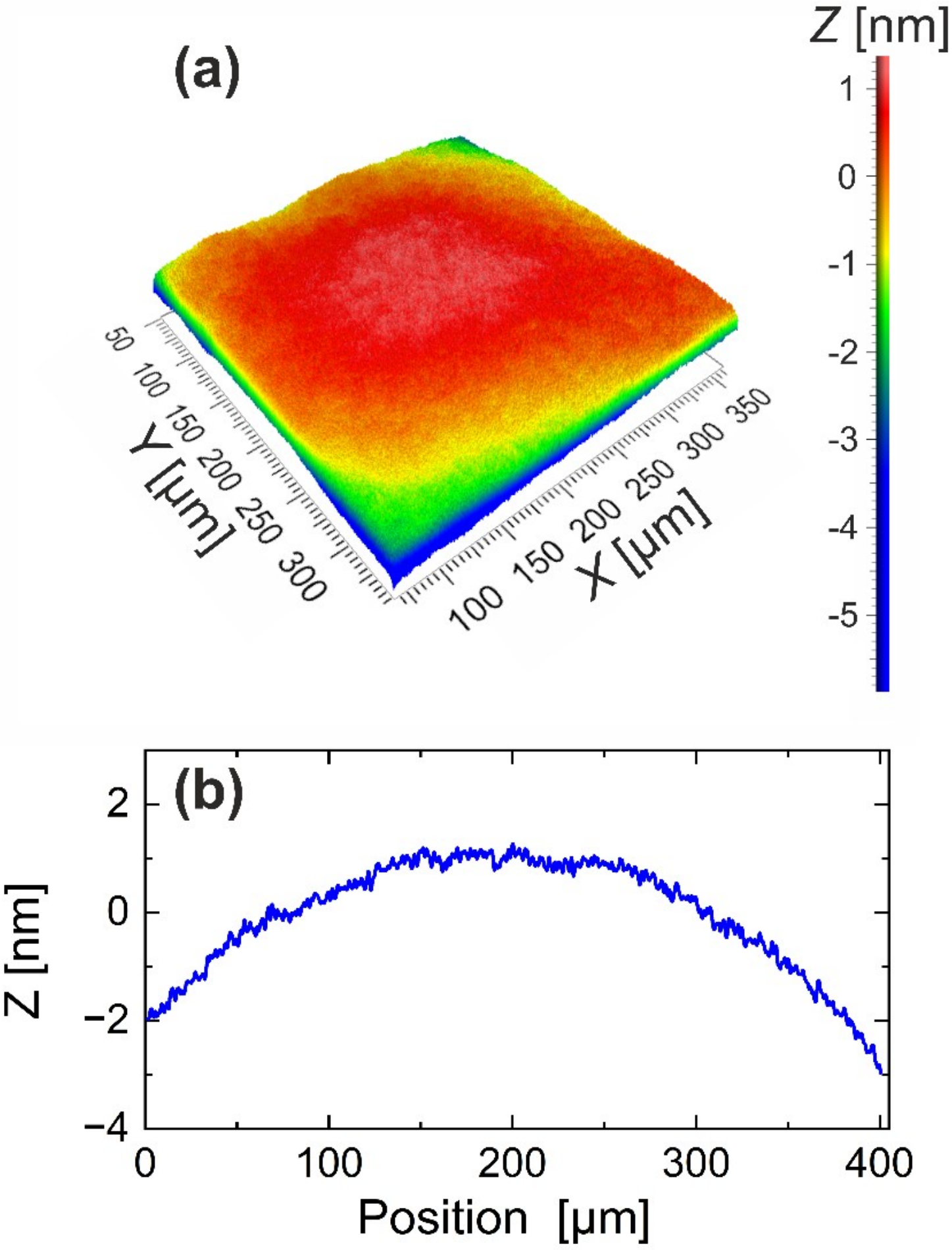

Figure 11: (a) Scanning white light interference microscope (SWLIM) topography of a 200 nm thick $SiO_2$ layer deposited by PVD on a 300 nm thick $Si_3N_4$ membrane window. (b) Cross-section along the diagonal.

Selected laser-irradiated surfaces were inspected at the EuXFEL by electron microscopy after overcoating the sample with a thin layer of gold for ensuring the necessary conductivity. For that, a scanning electron microscope (FEI Quanta FEG 650, Hillsboro, USA) was operated in secondary electron imaging mode, using an electron acceleration voltage of 15 kV.

### *5.2 Data Processing and Evaluation*

As a first step, a dark image acquired by the CCD camera with both, the laser and the X-ray beams blocked was subtracted from the experimentally recorded SAXS patterns. All images were normalized by a signal proportional to the corresponding X-ray pulse flux, compensating for fluctuations in the X-ray probe beam. In the pump-probe experimental runs, outliers were removed manually from the set of six images recorded under nominally identical experimental parameters (laser fluence, delay), before averaging. If applicable, scattering images recorded at negative delay times (for pump-probe delay scans), acquired with the laser beam blocked (for pump-probe fluence scans), or taken before the first laser pulse exposure (for in-situ pulse-by-pulse experiments) were subtracted from all images. This enabled background subtraction and to isolate the laser-induced changes. Moreover, in the pump-probe experiments, the total intensity scattered around the beam stop position was spatially integrated and used for additional normalization of the SAXS images, thereby compensating for fluctuations in the X-ray probe beam flux. Finally, the SAXS images were smoothed by a Gaussian filter that averaged over 2.4 pixels in both the $q_x$- and $q_y$-directions.

To quantitatively evaluate specific scattering features in the recorded SAXS patterns, data post-processing was performed. As the signatures of LIPSS features are oriented mainly parallel or perpendicular to the laser beam polarization and extend over a limited azimuthal angular range, in a first step, the scattered intensities $I_{SAXS}(q_x, q_y)$ were transformed into a polar coordinate system $I_{SAXS}(q_r, q_\varphi)$. In this representation, the data were integrated numerically in the azimuthal range of interest $I'_{SAXS}(q_r) = \int I_{SAXS}(q_r, q_\varphi)\, dq_\varphi$ to include all relevant scattering signals. For LIPSS scattering signatures azimuthal integration ranges $\Delta\varphi$ were chosen between = ± 20° and ± 25° around the polarization direction and the direction orthogonal to it. For ring-shaped scattering features, a full azimuthal integration ($\Delta\varphi = 360°$) was performed. In a third step, the derived profiles $I'_{SAXS}(q_r)$ were decomposed by least-squares-fits into an exponentially decaying background signal around the beam stop and a signal represented by skewed Gaussian peaks. The latter were then characterized by their peak position $q_{r,peak}$, amplitude $I_{Peak/R}$, area, skewness, as well as the full width at half maximum (FWHM) $\Delta q_{FWHM}$. The most frequent spatial period was calculated from the peak position via $\Lambda = 2\pi/q_{r,peak}$.

## Acknowledgements

The authors acknowledge European XFEL in Schenefeld, Germany, for provision of X-ray free-electron laser beam time at SCS SASE3 under proposal number 5697 and would like to thank the staff for their exceptional support. The access to the European XFEL was supported by a grant of the Polish Ministry of Education and Science - decision No. 2022/WK/13. D.K., T.J.A., and K.S.T. acknowledge support by the Deutsche Forschungsgemeinschaft (DFG, German Research Foundation) – Project No. 278162697-SFB 1242. J.B. acknowledges support by the Deutsche Forschungsgemeinschaft (DFG, German Research Foundation) – Project No. 530345255. M.M. acknowledges support by U.S. Department of Energy, Laboratory Directed Research and Development program at SLAC National Accelerator Laboratory, under contract DE-AC02-76SF00515.

Open access funding enabled and organized by Projekt DEAL.

**Author Contributions**

K.S.T. and J.B. conceptualized the research. G.M., M.T., R.C., K.S.T. and J.B. developed the experimental methodology. J.L.D. and J.B. developed the theoretical modelling methodologies. D.K., J.L.D., and T.J.A. developed the modelling software used in this research. M.W. prepared the $SiO_2$ film samples used in this study and performed the SWLIM characterization. J.L.D. performed the FDTD simulations and curated the results. D.K. and J.B. performed and the Sipe-Drude modelling and curated the results. All authors contributed to the investigation. All authors contributed to the validation of the data. D.K., T.J.A., M.W., K.S.T., and J.B. curated the experimental data. D.K. and T.J.A. performed the formal analysis of the experimental data. D.K. and J.B. visualized the data. J.B. prepared the original draft of the manuscript. D.K., T.J.A., K.S.T., and J.B. revised the original manuscript draft. All authors reviewed and edited the final manuscript. G.M. and K.S.T. administrated the project. K.S.T. and J.B. acquired funding for this research. K.S.T. and J.B. supervised the project.

**Conflict of interest**

The authors declare no competing interests.

**Data availability**

The original European XFEL data will be publicly available after an embargo period of three years (DOI: https://doi.org/10.22003/XFEL.EU-DATA-005697-00 .

**Supporting Information**

Supporting Information is available from the Wiley Online Library or from the authors.

## Supporting Information

# Probing the formation of femtosecond laser-induced periodic surface structures on silica films by ultrafast small-angle X-ray scattering

*Dominik Kaczmarek*[#]*, Jean-Luc Déziel, Thies Johannes Albert, Matthias Weise, Jerzy Antonowicz, Robert Carley, Devesh Chopra, Loïc Le Guyader, Laurent Mercadier*[❀]*, Giuseppe Mercurio, Roman Minikayev, Mianzhen Mo, Andreas Scherz, Ryszard Sobierajski, Yanwen Sun, Martin Teichmann, Peter Zalden, Klaus Sokolowski-Tinten, and Jörn Bonse*[*]

[#] dominik.kaczmarek@uni-due.de / * joern.bonse@bam.de

### Supplementary Text

S1. Fluence Dependence of Pulse-by-Pulse In-Situ SAXS of LIPSS Formation

In analogy to Figure 2 shown in *Section 3.1*, pulse-by-pulse in-situ SAXS measurements were performed at two additional peak laser fluences, $\phi_0$. **Figures S1 and S2** and present collages of selected SAXS patterns after exposure to different numbers of laser pulses, *N*, applied to the same sample spot, at a fixed peak fluence of $\phi_0$ = 0.30 J cm$^{-2}$. (Figure S1) or $\phi_0$ = 0.36 J cm$^{-2}$ (Figure S2), respectively.

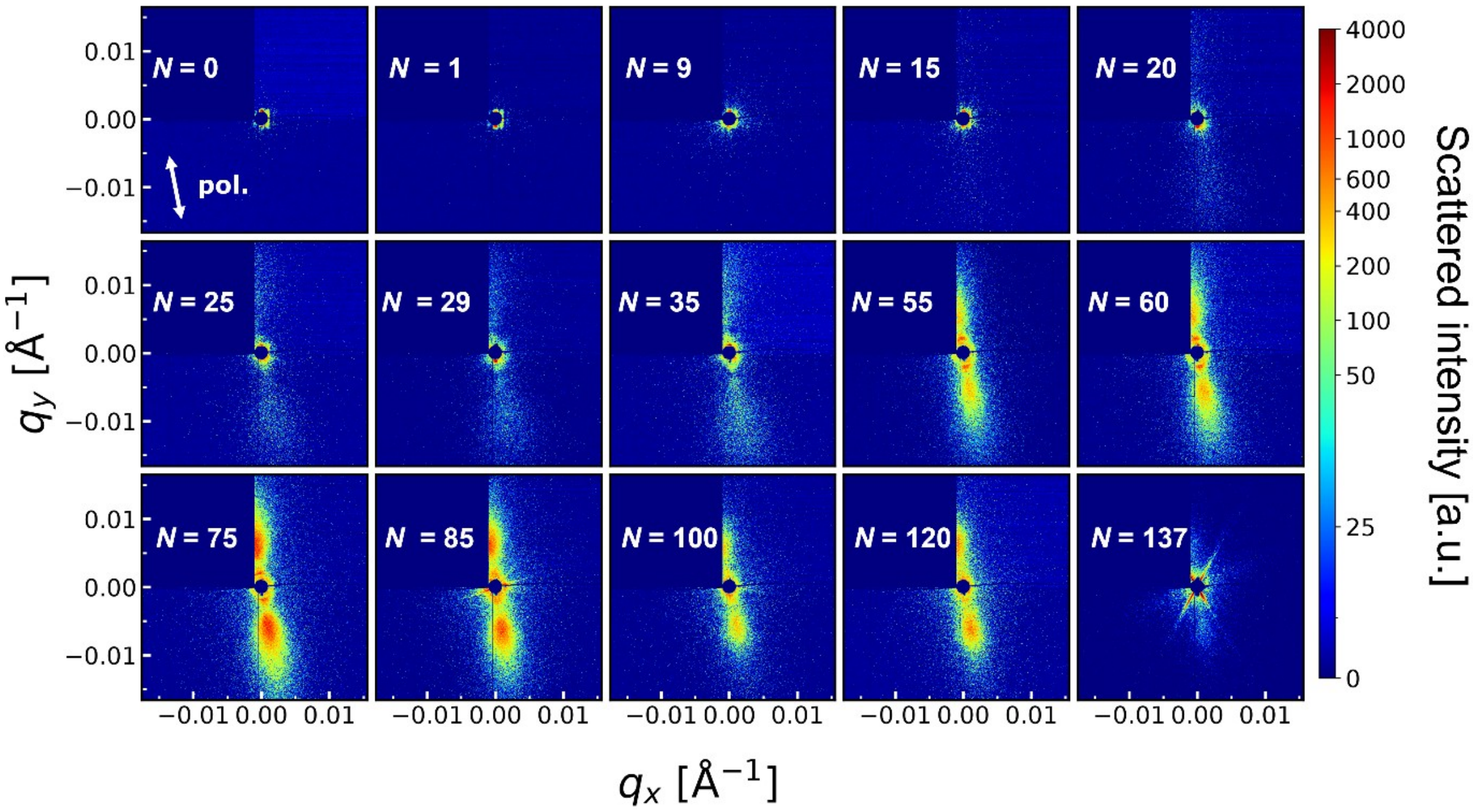

Figure S1: Selected SAXS patterns after exposure to different numbers of laser pulses *N,* applied to the same sample spot, at a fixed peak fluence of $\phi_0$ = 0.30 J cm$^{-2}$. All images are encoded in a joint color scale for the scattered intensity. The white double-arrow in the frame for *N* = 1 displays the projection of the direction of the linear laser beam polarization onto the sample surface.

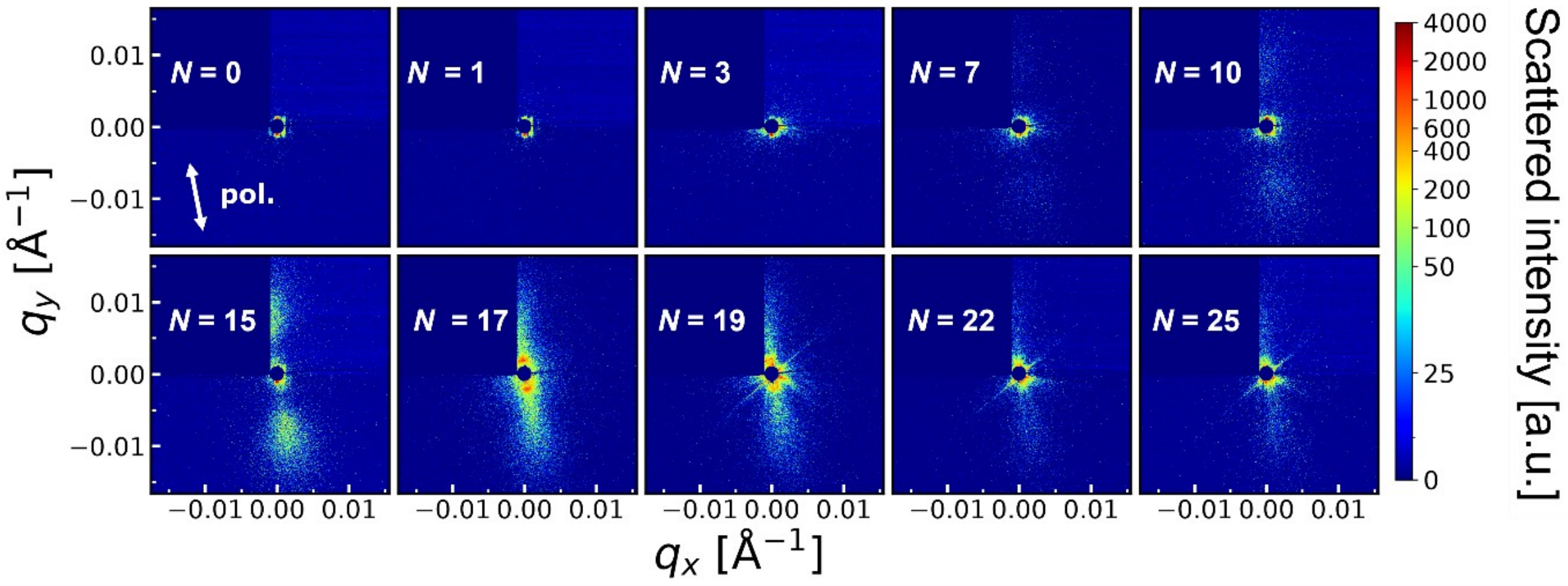

Figure S2: Selected SAXS patterns after exposure to different numbers of laser pulses *N*, applied to the same sample spot, at a fixed peak fluence of $\phi_0$ = 0.36 J cm$^{-2}$. All images are encoded in a joint color scale for the scattered intensity. The white double-arrow in the frame for *N* = 1 displays the projection of the direction of the linear laser beam polarization onto the sample surface.

The separately available MP4-file S3 animates the full series of 274 SAXS patterns, recorded pulse-by-pulse at a peak laser fluence of $\phi_0$ = 0.27 J cm$^{-2}$ at the same sample location, in a video. Selected SAXS patterns of the series are displayed in Figure 2 in Sect. 3.1. The apparent discontinuous transitions represent discrete film delamination/ablation events.

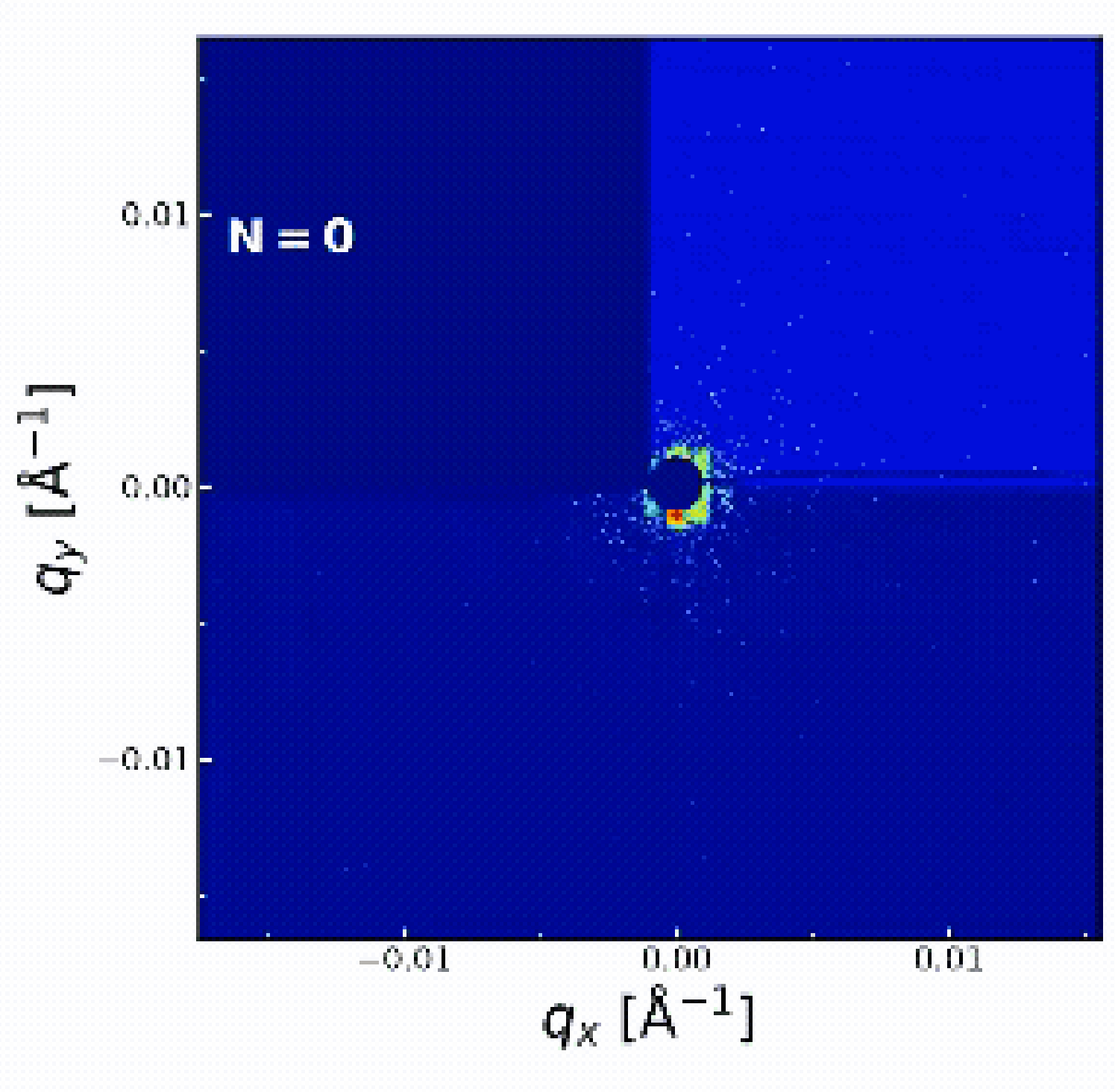

Movie S3: Motion picture collage of all SAXS patterns after exposure to different numbers of laser pulses *N*, applied to the same sample spot, at a fixed peak fluence of $\phi_0$ = 0.27 J cm$^{-2}$.

All images are encoded in a joint color scale for the scattered intensity. A higher resolution MP4-movie is available as separate Supporting Information file.

## S2. Fluence Dependence of Single-Pulse Pump-Probe SAXS Pattern Characteristics

In analogy to Figure 6(b) shown in Sect. 3.2.2, the characteristic features of the SAXS patterns presented in Figure 7 were evaluated by azimuthal integration and least-squares-fitting procedures to obtain the nanoscatterers and HSFL-I characteristics as function of the peak laser fluence $\phi_0$ at a fixed pump-probe delay time $\Delta t$ = +500 ps.

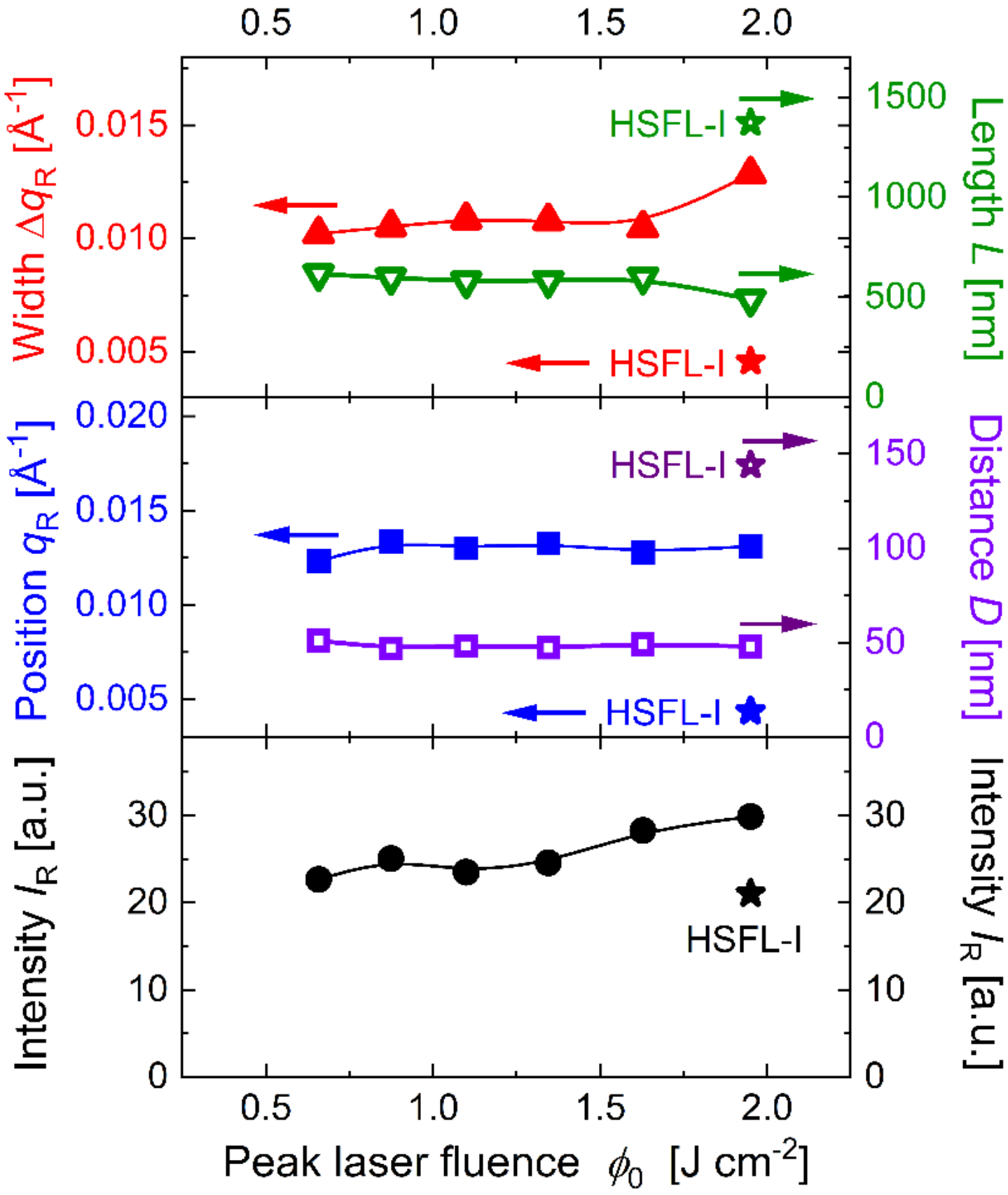

Figure S4: Nanoscatterer characteristics as function of the peak laser fluence $\phi_0$ at a fixed pump-probe delay time $\Delta t$ = +500 ps. Left abscissa: ring width (FWHM) $\Delta q_R$ (top panel), peak position $q_R$ (middle panel), and peak scattering intensity $I_R$ (bottom panel), as derived by azimuthal integration over the SAXS patterns presented in Figure 7; Right abscissa: coherent scattering length $L$ (top panel) and correlation distance of scattering centers $D$ (middle panel). The star-shaped data points represent the characteristics of the HSFL-I signatures. The lines guide the eye.